\documentclass[osajnl,twocolumn,showpacs,superscriptaddress,10pt]{revtex4-1} 
\usepackage{amsmath,amssymb,graphicx}
\usepackage{caption}
\usepackage{graphicx}
\usepackage{color}
\usepackage[justification=RaggedRight, singlelinecheck=false]{caption}
\usepackage{ulem}

\begin{document}

\title{Polarization dependence of nonlinear wave mixing of spinor polaritons in semiconductor microcavities}

\author{Przemyslaw Lewandowski}
\affiliation{Physics Department and Center for Optoelectronics and Photonics Paderborn (CeOPP), Universit\"at Paderborn, Warburger Strasse 100, 33098 Paderborn, Germany}

\author{Ombline Lafont}
\affiliation{Laboratoire Pierre Aigrain, \'Ecole Normale Sup\'erieure - PSL Research University, CNRS, Universit\'e Pierre et Marie Curie - Sorbonne Universit\'es, Universit\'e Paris Diderot - Sorbonne Paris Cit\'e, FR-75231 Paris Cedex 05, France}

\author{Emmanuel Baudin}
\affiliation{Laboratoire Pierre Aigrain, \'Ecole Normale Sup\'erieure - PSL Research University, CNRS, Universit\'e Pierre et Marie Curie - Sorbonne Universit\'es, Universit\'e Paris Diderot - Sorbonne Paris Cit\'e, FR-75231 Paris Cedex 05, France}

\author{Chris K. P. Chan}
\affiliation{Department of Physics, The Chinese University of Hong Kong, Hong Kong SAR, China}

\author{P. T. Leung}
\affiliation{Department of Physics, The Chinese University of Hong Kong, Hong Kong SAR, China}
\affiliation{Center of Optical Sciences, The Chinese University of Hong Kong, Hong Kong SAR, China}

\author{Samuel M. H. Luk}
\affiliation{College of Optical Sciences, University of Arizona, Tucson, AZ 85721, USA}
\affiliation{Department of Physics, University of Arizona, Tucson, AZ 85721, USA}
 
\author{Elisabeth Galopin}
\affiliation{Laboratoire de Photonique et de Nanostructures, Centre National de la Recherche Scientifique, Route de Nozay, FR-91460 Marcoussis, France}
\author{Aristide Lema\^itre}
\affiliation{Laboratoire de Photonique et de Nanostructures, Centre National de la Recherche Scientifique, Route de Nozay, FR-91460 Marcoussis, France}
\author{Jacqueline Bloch} 
\affiliation{Laboratoire de Photonique et de Nanostructures, Centre National de la Recherche Scientifique, Route de Nozay, FR-91460 Marcoussis, France}

\author{Jerome Tignon}
\affiliation{Laboratoire Pierre Aigrain, \'Ecole Normale Sup\'erieure - PSL Research University, CNRS, Universit\'e Pierre et Marie Curie - Sorbonne Universit\'es, Universit\'e Paris Diderot - Sorbonne Paris Cit\'e, FR-75231 Paris Cedex 05, France}

\author{Philippe Roussignol}
\affiliation{Laboratoire Pierre Aigrain, \'Ecole Normale Sup\'erieure - PSL Research University, CNRS, Universit\'e Pierre et Marie Curie - Sorbonne Universit\'es, Universit\'e Paris Diderot - Sorbonne Paris Cit\'e, FR-75231 Paris Cedex 05, France}

\author{N. H. Kwong}
\affiliation{College of Optical Sciences, University of Arizona, Tucson, AZ 85721, USA}

\author{Rolf Binder}
\affiliation{College of Optical Sciences, University of Arizona, Tucson, AZ 85721, USA}
\affiliation{Department of Physics, University of Arizona, Tucson, AZ 85721, USA}

\author{Stefan Schumacher}
\affiliation{Physics Department and Center for Optoelectronics and Photonics Paderborn (CeOPP), Universit\"at Paderborn, Warburger Strasse 100, 33098 Paderborn, Germany}
\affiliation{College of Optical Sciences, University of Arizona, Tucson, AZ 85721, USA}

\begin{abstract}

The pseudo-spin dynamics of propagating exciton-polaritons in semiconductor microcavities are known to be strongly influenced by TE-TM splitting. As a vivid consequence, in the Rayleigh scattering regime, the TE-TM splitting gives rise to the optical spin Hall effect (OSHE). Much less is known about its role in the nonlinear optical regime in which four-wave mixing for example allows the formation of spatial patterns in the polariton density, such that hexagons and two-spot patterns are observable in the far field. 
Here we present a detailed analysis of spin-dependent four-wave mixing processes, by combining the (linear) physics of TE-TM splitting with spin-dependent nonlinear processes, i.e., exciton-exciton interaction and fermionic phase-space filling.
Our combined theoretical and experimental study elucidates the complex physics of
the four-wave mixing processes that govern polarization and orientation of off-axis modes. 
\end{abstract}

\maketitle

\section{Introduction}  
\label{sec:intro}
Over the past decades exciton-polaritons in planar semiconductor microcavities have been intensively investigated both for their fundamental interest and, more recently, also for their potential applications \cite{Balarini2013,dawes.09}.
Many of the intriguing features of cavity polaritons are associated with their particular energy-angle dispersion. 
A characteristic of this dispersion is the small effective mass of the polaritons which has allowed for several important developments
 such as the successful observation of polariton Bose-Einstein condensates
\cite{deng-etal.02,Kasprzak2006,Balili2007,utsunomiya-etal.08,deng-etal.10,snoke-littlewood.10,moskalenko-snoke.00}, or new spectroscopic insights into the Coulomb interaction between excitons \cite{gonokami-etal.97,fan-etal.98,baars-etal.00,kwong-etal.01prl,kwong-etal.01prb}.
Another characteristic of the energy-angle dispersion is an inflection point in the lower polariton branch called the “magic angle”.
When a strong beam pumps the polaritons at this angle, a weak probe beam in normal incidence experiences large parametric amplification \cite{Savvidis2000,Huang2000,Ciuti2000,Stevenson2000,houdre-etal.00,Saba2001,Whittaker2001,%
Ciuti2003,Savasta2003,Savasta2003b,langbein.04,Baumberg2005,Keeling2007,sanvitto-timofeev.12}.
More recently, applications of the exciton-polariton physics such as all-optical switching and all-optical transistors have been discussed
\cite{dawes.09,schumacher-etal.09pssrrl,dawes-etal.10,Balarini2013,Schmutzler2014}.\\

Polariton-polariton scattering, as a consequence of the Coulomb interaction between the underlying excitons, can lead to wave mixing processes such as four-wave mixing (FWM). These are utilized in stimulated scattering and parametric amplification, and in certain geometries they can exhibit threshold behavior as a function of polariton density: wave mixing leads to optical instability and possibly becomes self-sustained above the optical parametric oscillation threshold (OPO) \cite{Ardizzone2013}. But even in the linear optical regime, where polariton-polariton scattering is negligible, interesting dynamical processes have been found. These include the so-called optical spin Hall effect \cite{Kavokin2005,Leyder2007}, which is a consequence of  the TE-TM splitting in the microcavity. It involves only resonant Rayleigh scattering which gives rise to an occupation of the so-called elastic circle since it changes the wave vector but not the energy of the polaritons.
The TE-TM splitting lifts the two-fold polarization degeneracy between the transverse electric (TE) and transverse magnetic (TM) modes, and the resulting polariton dynamics have been cast into the form of torque-like pseudo-spin dynamics \cite{Kavokin2005,Leyder2007}. These effects related to TE-TM splitting, which can be described in a spinor formulation of the two-component polaritons, also affect nonlinear wave mixing \cite{Schumacher2007a,Schumacher2007c}.

\begin{figure} 
\includegraphics[scale=0.7]{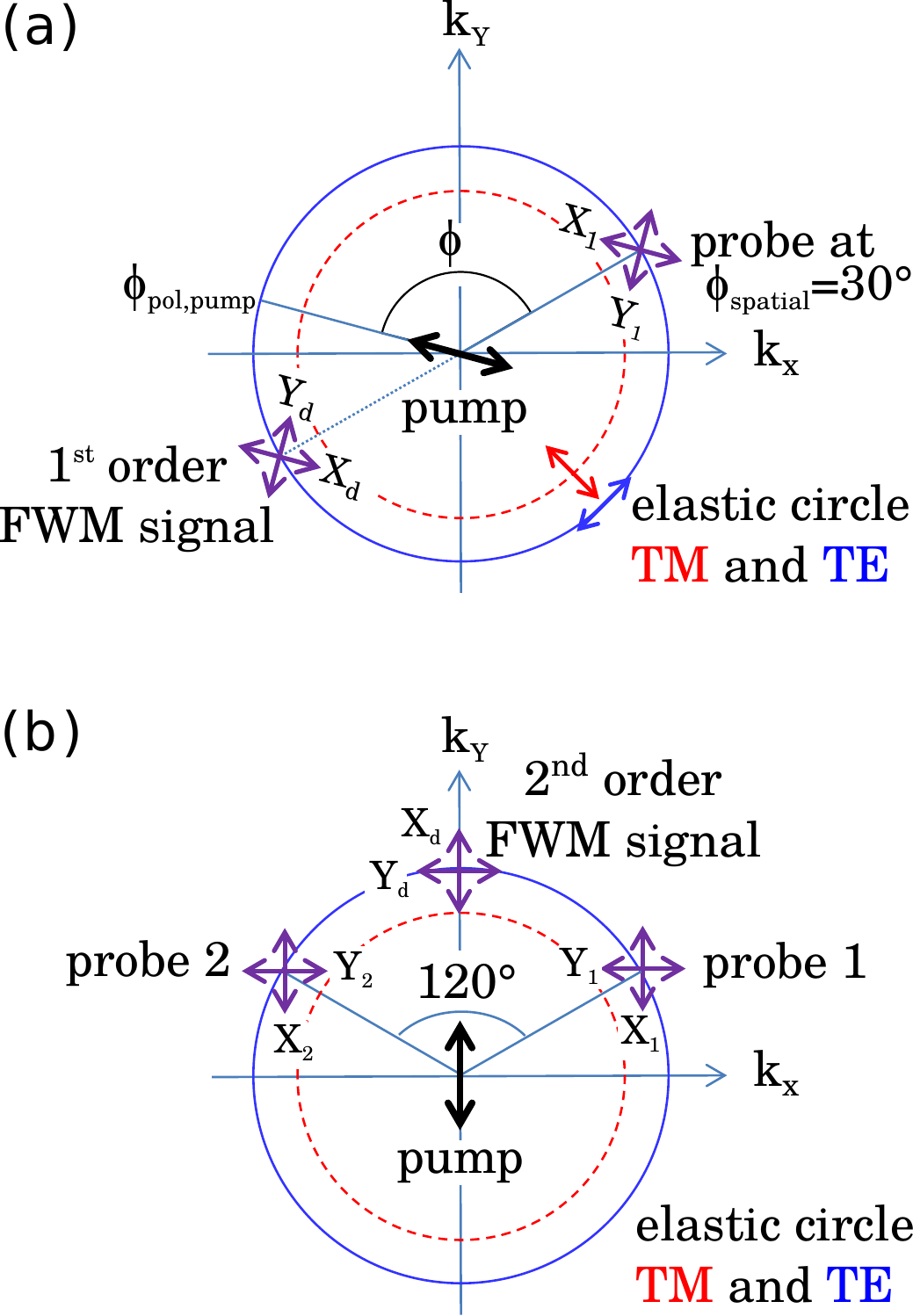} 
\captionof{figure}{Sketch of the two-dimensional transverse momentum space (k-space) plane and linear polarization states indicated by double arrows. The vector $(k_x,k_y)$ defines the emission direction in the far field. The TE and TM elastic circles of LPB$_1$ are indicated, with TE (TM) polarizations being tangential (orthogonal) to the elastic circle. The pump is in normal incidence at $k=0$ and linearly polarized.
In (a) the configuration used to study the FOFWM is illustrated.
Probe and FOFWM signals are at $\mathbf{k}_\mathrm{probe}$  corresponding to $\phi_{\mathrm{spatial}}=30^\circ$ and $-\mathbf{k}_\mathrm{probe}$ corresponding to $\phi_\mathrm{spatial}+180^\circ$, respectively. The pump polarization plane $\phi_\mathrm{pol,pump}$ is rotated during the experiment.
In (b), the configuration used to study the SOFWM is illustrated. The azimuthal angle $\phi_{\mathrm{spatial}}=30^\circ$ of probe 1 is fixed, and the relative angle between probe 1 and probe 2 is 120$^\circ$, leading to a SOFWM signal at 90$^\circ$. }
\label{fig:pump_probe_scheme}
\end{figure}

Nonlinear polariton wave mixing above the OPO threshold has recently been shown to result in spatial polariton patterns \cite{Ardizzone2013}. Theoretical investigations of these patterns
 \cite{Kheradmand2008,Luk2013,Egorov2014} include the formation of hexagonal patterns in the polariton density inside the cavity, which appear as six-spot (or hexagon) emission in the cavity's far field. A detailed analysis has revealed that two main FWM processes are crucial for the hexagon formation process. On the one hand, scattering states separated by 180$^\circ$ on the elastic circle (Fig.~\ref{fig:pump_probe_scheme}a) involve two pump polaritons (at zero in-plane momentum) and one probe polariton (anywhere on the elastic circle). We will refer to this process as first-order FWM (FOFWM).
On the other hand, states separated by 60$^\circ$ on the elastic circle involve one pump polariton and two probe polaritons, the latter being separated by 120$^\circ$ on the elastic circle. We will refer to it as the second-order FWM (SOFWM, illustrated in Fig.~\ref{fig:pump_probe_scheme}b).
 
In this paper, we explore the underlying physics of these FWM processes in light of the non-trivial effects due to TE-TM splitting, and study both experimentally and theoretically their polarization dependence. We operate in the nonlinear optical regime, but slightly below the OPO threshold so that the spatially homogenous fields are still stable.

\begin{figure}
\includegraphics[scale=0.35]{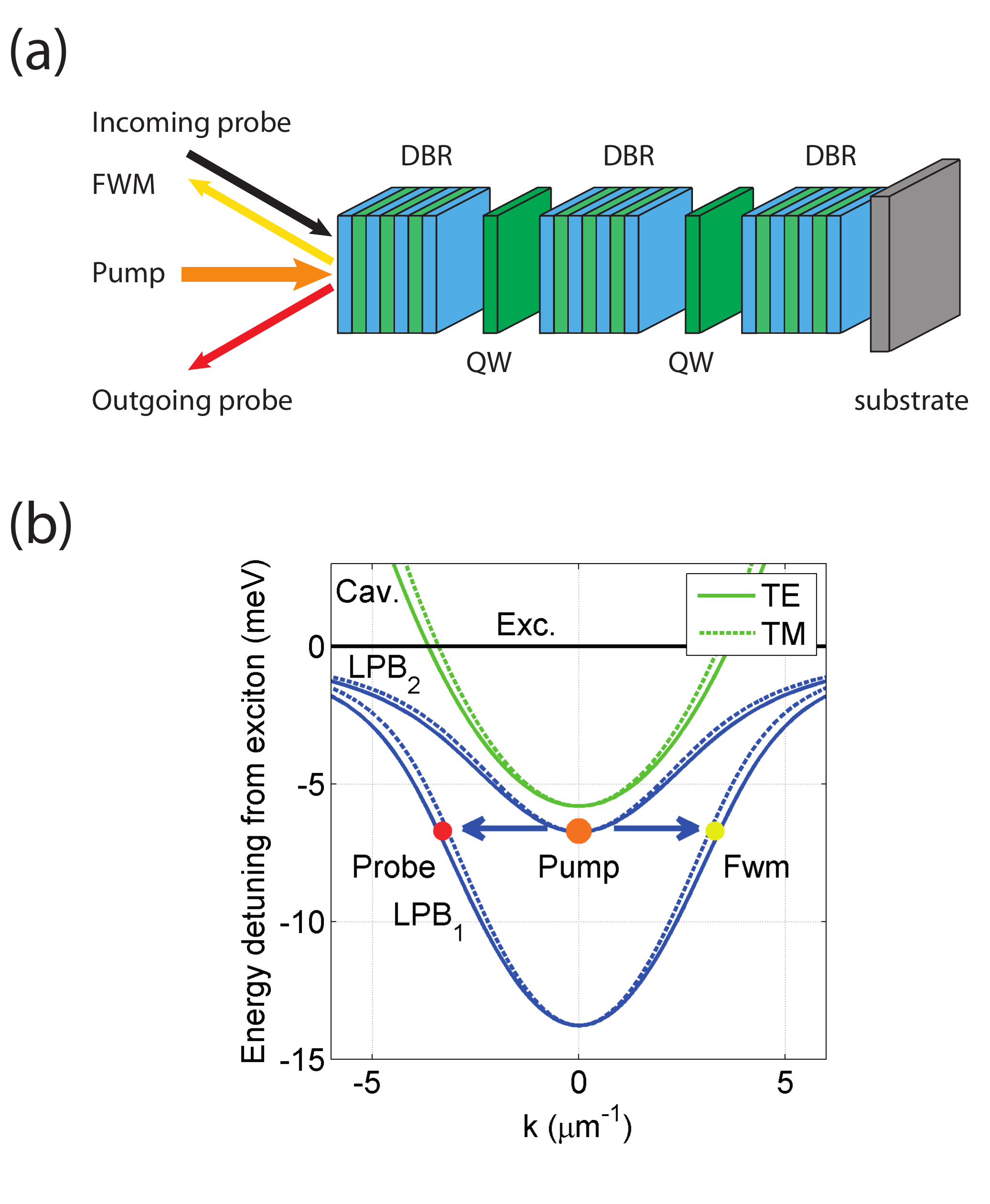}
\caption{(a) Sketch of the double microcavity studied and the excitation-detection geometry.  Applying a continuous-wave pump and a continuous-wave probe, the first-order four-wave mixing signal is detected in reflection geometry.
(b) Shown are the polariton branches in the double cavity system. The bare cavity dispersion (green) and the bare exciton dispersion (black) split into four polariton-branches due to the strong inter-cavity coupling and the coupling between excitons and photons in each cavity. Only the two lower polariton branches LPB$_1$ and LPB$_2$ are shown. Splitting of the polariton branches into TE (solid lines) and TM (dashed lines) modes is included.}
\label{fig:dc_and_dispersion}
\end{figure}

The experimental sample, which consists of quantum wells embedded in two coupled cavities, is sketched in Fig \ref{fig:dc_and_dispersion}a. Each cavity is formed of distributed Bragg reflectors (DBR). A detailed description of the sample can be found in ref.~\cite{Ardizzone2013}. Fig.~\ref{fig:dc_and_dispersion}b shows the dispersion branches of this coupled system: the cavity exhibits an approximately quadratic dispersion due to the confinement in two dimensions (the cavity plane). The excitons, optically excited in the embedded quantum wells, have an approximately flat dispersion (due to the heavy effective exciton mass). The mutual coupling between cavity-photons and excitons and the cavity-cavity coupling give rise to a normal mode splitting in two upper polariton-branches and two lower polariton-branches, LPB$_1$ and LPB$_2$.

Fig.~\ref{fig:dc_and_dispersion}a illustrates the optical pump-probe setup used to study the FOFWM: a cw-pump excites the cavity at normal incidence ($\bold{k}_{\mathrm{pump}}=\bold{0}$) resonantly on LPB$_2$ and gives rise to a coherent polariton field. A cw-probe beam with the same frequency is applied under oblique incidence, resonant on LPB$_1$, carrying an in-plane momentum $\bold{k}_{\mathrm{probe}}$. For a sufficiently intense pump, the probe beam at $\bold{k}_{\mathrm{probe}}$ can stimulate a pairwise scattering of pump polaritons into off-axis modes. The two pump polaritons scatter phase-matched and resonant into the modes $\bold{k}_{\mathrm{probe}}$  - amplifying the probe - and $\bold{k}_{\mathrm{FOFWM}} = 2 \bold{k}_{\mathrm{pump}}-\bold{k}_{\mathrm{probe}} = -\bold{k}_{\mathrm{probe}}$, triggering a FOFWM signal.
For the SOFWM, a second cw probe with same frequency and polar angle of incidence is sent on the cavity (not shown in Fig.~\ref{fig:dc_and_dispersion}a). If the azimuthal angle separating the two probes is 120$^\circ$, phase matching conditions are fulfilled and the two probes can stimulate scattering of one pump polariton into an off-axis mode at $\bold{k}_{\mathrm{SOFWM}} = \bold{k}_{\mathrm{pump}}-\bold{k}_{\mathrm{probe1}}-\bold{k}_{\mathrm{probe2}}$, triggering a SOFWM signal.

Fig.~\ref{fig:pump_probe_scheme} illustrates the physical quantities of interest in the present study. In the case of FOFWM (Fig.~\ref{fig:pump_probe_scheme}a) $\phi_\mathrm{spatial}$ and $\phi_\mathrm{pol,pump}$ denote the azimuthal angle of the probe incidence plane and of the polarization plane of a linearly polarized pump, respectively. The relevant quantity for this study, however, is the relative angle between both, $\phi = \phi_{\mathrm{pol,pump}}-\phi_{\mathrm{spatial}}$.
The linearly polarized probe is either co- ($X_1$) or cross-polarized ($Y_1$) to the pump polarization and we detect the FOFWM signal decomposed in its co- ($X_d$) and cross-polarized ($Y_d$) part. 
In the case of SOFWM (Fig.~\ref{fig:pump_probe_scheme}b), we add a second probe separated by 120$^\circ$ with respect to the first probe, leading to a SOFWM signal at 90$^\circ$. The pump is vertically polarized and the probes polarizations can be either vertically polarized (i.e. X, copolarized to the pump's polarization) or horizontally polarized (i.e. Y, cross-polarized to the pump's polarization). Again, we measure the polarization of the resulting SOFWM signal.

The paper is organized as follows.
In Sec.~\ref{sec:single-probe-results} we present results of the FOFWM experiment and compare them to theoretical results  obtained from a numerical solution of a general theoretical model for the light amplitude and excitonic polarization densities. In order to analyze and understand the experimental and numerical results, we discuss in Sec.~\ref{sec:sec:linear-optical-regime} the  polarization dependence in the linear optical regime, followed by a study of the nonlinear response at the level of a linear stability analysis in Sec.~\ref{sec:sec:LSA} and a discussion in Sec.~3.C.
In Sec.~\ref{sec:sec:double_probe} we present results and analysis for the SOFWM experiment, and conclude with
Sec.~\ref{sec:conclusion}.
\section{First Order Four-Wave Mixing}
\label{sec:single-probe-results}
First we focus on those four-wave mixing processes that can be interpreted as off-axis scattering of two pump-induced polaritons to finite and opposite momentum $\mathbf{k}$ and $-\mathbf{k}$ onto the elastic circle as indicated in Fig.~\ref{fig:pump_probe_scheme}a.
In a four-wave mixing context, the dominant contribution to this process is only of linear order in the off-axis probe intensity.
To selectively probe these processes, we perform a polarization resolved pump and probe experiment as explained in Sec. 1.  We fix the probe azimuthal angle $\phi_\mathrm{spatial}=30^\circ$ and rotate the polarization plane of the linearly polarized pump $\phi_\mathrm{pol,pump}$ stepwise. For each $\phi_\mathrm{pol,pump}$, the linear probe is re-adjusted either copolarized ($X$) or cross-polarized ($Y$) to the pump. We then measure the intensity of the FOFWM signal in $X$ and $Y$ polarization channel. The pump and probe intensities are fixed during each measurement.

The pump beam used is produced by a Ti:Saphir laser operated at 775~nm with an intensity of $100\,\mathrm{mW}$. All experiments are performed at $6\,\mathrm{K}$ using a coldfinger cryostat.
The pump and probe spots on the sample are around $50\,\mathrm{\mu m}$ full-width half maximum. 

Our theoretical analysis is based on a microscopic description of the coupled nonlinear photon and exciton dynamics inside the double-cavity system as introduced in Ref.~\cite{Ardizzone2013}. The equation of motion for the cavity-field $E$ is based on classical light-theory in quasi-mode approximation. The dynamics of the excitonic polarization $p$ is derived within a density matrix theory approach in the coherent limit consistently taking into account all third order nonlinearities in 1s exciton approximation \cite{Luk2013}. First we formulate the theory in the circular polarization basis where the nonlinear interactions take their most simple form. The relevant equations of motion are then given by (note that the dependence of the dynamical fields on $\mathbf{r}$ and $t$ is suppressed for clarity):
\begin{align}
\mathrm{i}\hbar\dot{E}_{i}^{\pm} = & \left(\mathbb{H}_{c}-\mathrm{i}\gamma_{c}\right)E_{i}^{\pm} \nonumber \\ &+\mathbb{H}^{\pm} E_{i}^{\mp}-\Omega_{c}E_{j}^{\pm}-\Omega_{x}p_{i}^{\pm}+E_{\mathrm{pump},i}^{\pm} \label{eq:GPE_e}\, ,\\
\mathrm{i}\hbar\dot{p}_{i}^{\pm} = & \left(\mathbb{H}_{x}-\mathrm{i}\gamma_{x}\right)p_{i}^{\pm}-\Omega_{x}\left(1-\alpha_\mathrm{PSF}\vert p_{i}^{\pm}\vert^{2}\right)E_{i}^{\pm} \nonumber \\ &+T^{++}\vert p_{i}^{\pm}\vert^{2}p_{i}^{\pm}+T^{+-}\vert p_{i}^{\mp}\vert^{2}p_{i}^{\pm}.
\label{eq:GPE_p}
\end{align}
Here, $i \neq j$ is the cavity index, $1$ or $2$, and $\pm$ denotes the circular polarization states. The operators $\mathbb{H}_{c}$ and $\mathbb{H}_{x}$ include the dispersions of photon modes and exciton and $\gamma_c=0.05\,\mathrm{meV}$ and $\gamma_x=0.05\,\mathrm{meV}$ phenomenologically describe loss of photons from the cavity and excitonic decoherence, respectively. The exciton is considered to be dispersionless with $\mathbb{H}_{x}=\varepsilon^x_0=\mathrm{const}$. The photonic dispersion is given by $\mathbb{H}_{c}=\hbar\omega^c_0-\frac{\hbar^2}{4} \left( \frac{1}{m_\mathrm{TE}}+\frac{1}{m_\mathrm{TM}}\right) \left( \frac{\partial^2}{\partial x^2} +  \frac{\partial^2}{\partial y^2}\right)$, where $\hbar\omega^c_0=E_0^x-5.8\mathrm{meV}$ is the energy of the photonic ground state and $m_\mathrm{TM}=0.226\,\mathrm{meV ps^2\mu m^{-2}}$ and $m_\mathrm{TE}=0.236\, \mathrm{meV ps^2\mu m^{-2}}$ are the effective masses used for the parabolic TE and TM cavity mode dispersions, respectively. In the circular polarization basis,
TE-TM splitting leads to a coupling between $+$ and $-$ components in the photonic part, included by the operator $\mathbb{H}^{\pm} = \frac{\hbar^2}{4} \left( \frac{1}{m_\mathrm{TE}}-\frac{1}{m_\mathrm{TM}}\right) \left( \frac{\partial}{\partial x} \mp \mathrm{i}\frac{\partial}{\partial y}\right)^2$.
The coupling between both cavities is included by $\Omega_{c}=5.05\,\mathrm{meV}$ and the coupling between photons and excitons
in each cavity by $ \Omega_{x}=6.35\,\mathrm{meV}$. Both coupling constants are considered to be real valued.
TE-TM splitting strength, energy detuning between excitonic and photonic ground state, and the coupling constants $\Omega_c$ and $\Omega_x$ are experimentally determined for the specific sample.

The nonlinearities included through the excitonic part are: (i) phase-space filling
(PSF) with magnitude $\alpha_\mathrm{PSF}=5.188\cdot 10^{-4}\,\mathrm{\mu m^2}$ \cite{Luk2013}, (ii) the repulsive interaction between co-circularly polarized
excitons $T^{++}= 5.69 \cdot 10^{-3}\,\mathrm{meV \mu m^2}$, (iii) and an attractive interaction between counter-circularly polarized excitons
$T^{+-}= -T^{++}/3$ \cite{Ardizzone2013}. Any time retardation in the interaction is neglected which is well justified for the almost monochromatic excitation considered here. As in the experiments, the cavity field is pumped with a monochromatic continuous wave source at $k=0$, $E_{\mathrm{pump},i}^{\pm}=E_{\mathrm{pump},i}^{\pm}(\bold{r}) \exp(-\mathrm{i}\omega_{\mathrm{pump}} t)$. Here, $\omega_{\mathrm{pump}}$ is the pump frequency and $E_{\mathrm{pump},i}^{\pm}(\bold{r}) $ is the spatial pump profile. 
From a nonlinear transfer matrix calculation of the optical modes inside the double-cavity system, for the $\mathrm{LPB_2}$ energy minimum, we find a ratio of electric field amplitudes at the quantum wells in each of the two cavities of $E^\pm_1/E^\pm_2 = -1.53$. The change in sign reflects the partial anti-symmetry of this cavity mode. In the quasi-mode approach used in the present paper, we choose the source fields in each cavity, $E_{\mathrm{pump},1}^{\pm}$ and $E_{\mathrm{pump},2}^{\pm}$, such that the correct cavity-field ratio is obtained in the calculated steady state solution. We note that the exact ratio sensitively depends on the cavity design and therefore on the specific position on the sample in the experiments.

At a pump frequency detuning of $\hbar \omega_{\mathrm{pump}}-\varepsilon^x_0= -6.7\,\mathrm{meV}$, the pump is resonant with the minimum of LPB2. This allows an efficient elastic scattering on LPB1 at $k\approx 3.3\,\mathrm{\mu m^{-1}}$.
To simulate a pump-probe scenario, a weaker cw probe beam at the pump frequency and with a finite momentum is added in Eq.~\ref{eq:GPE_e}. We solve Eqs.~(\ref{eq:GPE_e}) and (\ref{eq:GPE_p}) explicitly in time on a finite sized Cartesian grid in real-space using a 4th order Runge-Kutta algorithm until signals in probe and FOFWM direction are stationary.

\begin{figure}
\begin{tabular}{@{}cc@{}cc@{}}
Experiment & Simulation \\
\includegraphics[height=3cm]{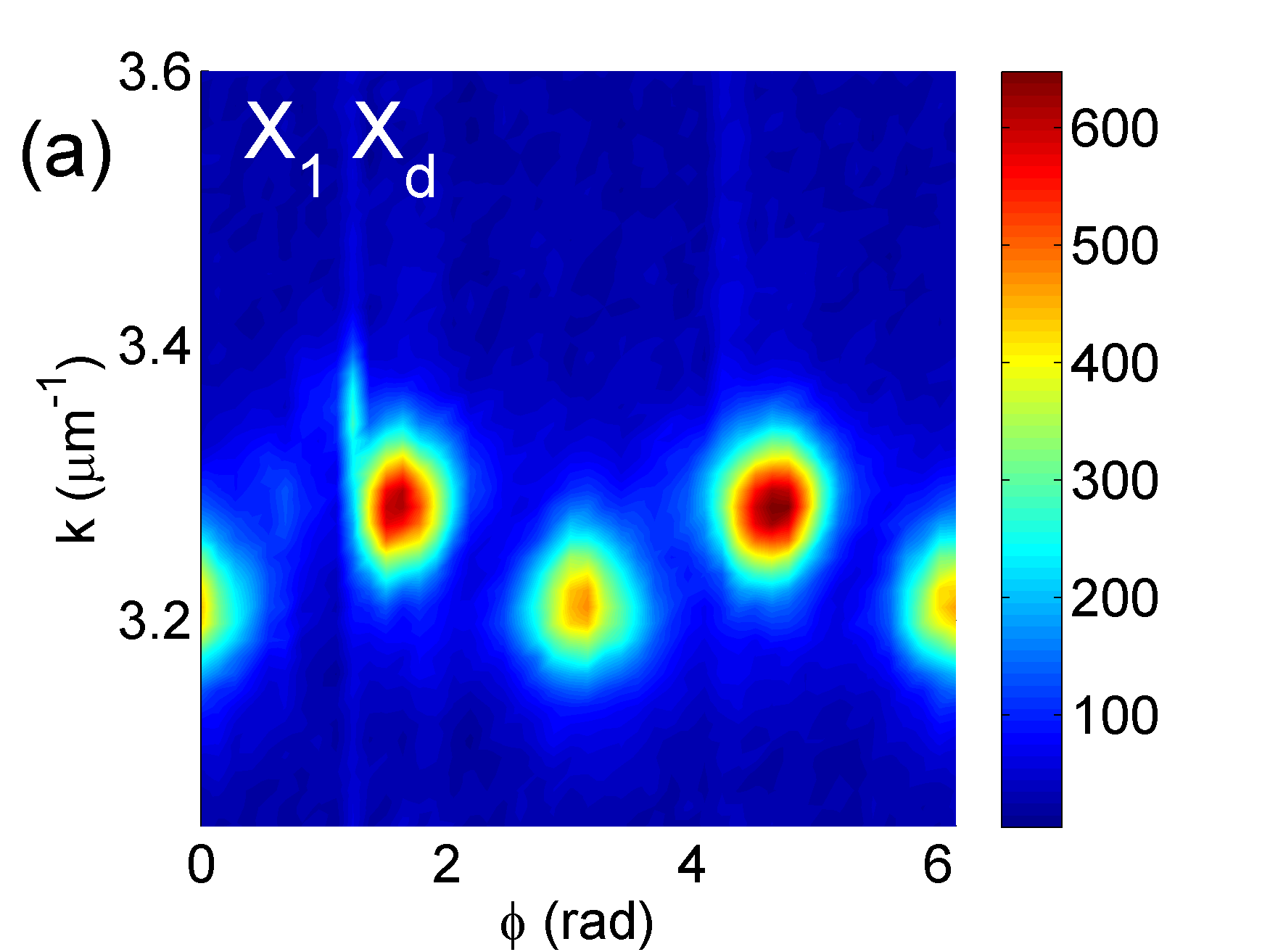}  &  \includegraphics[height=3cm]{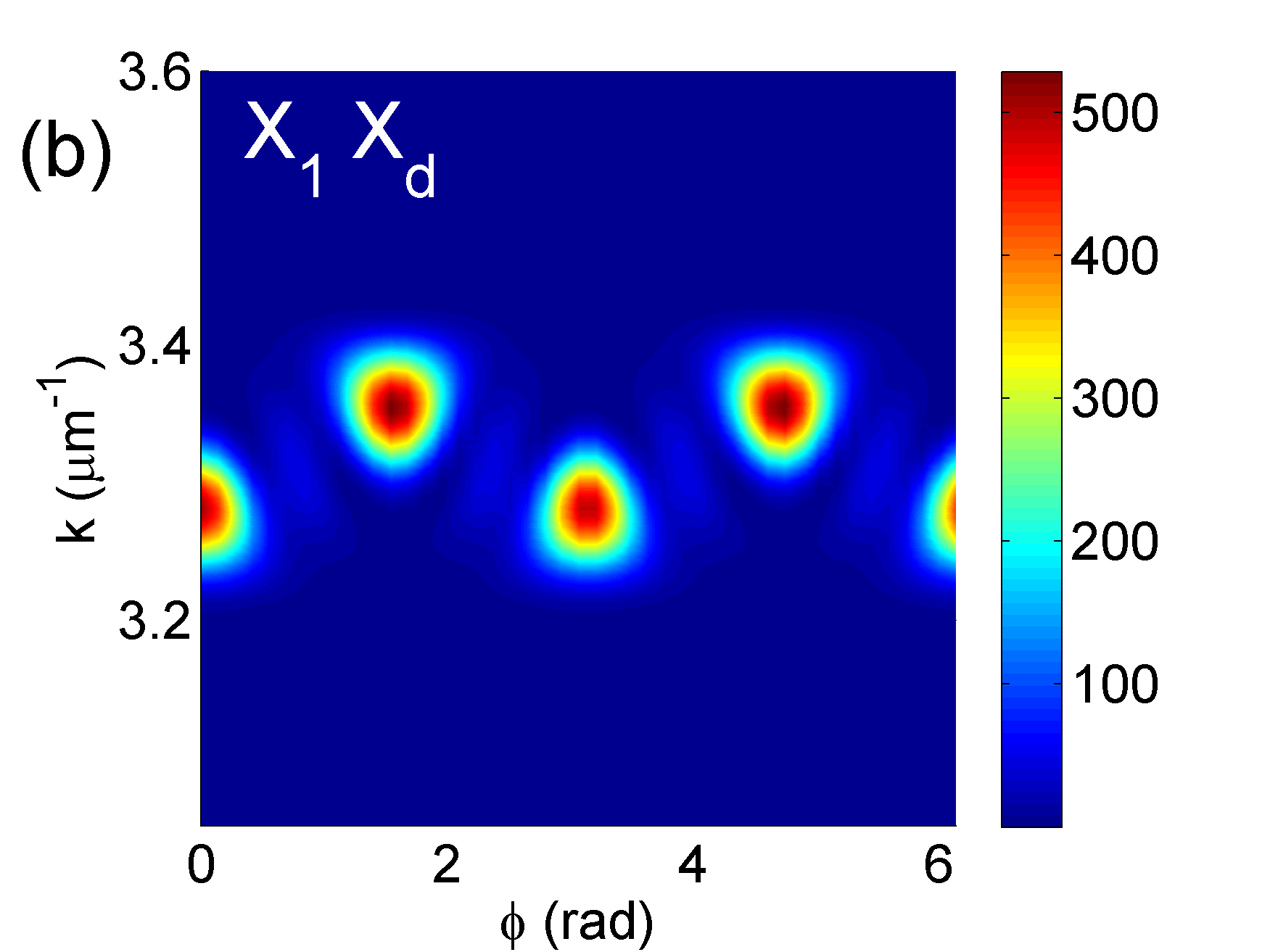}\\
\includegraphics[height=3cm]{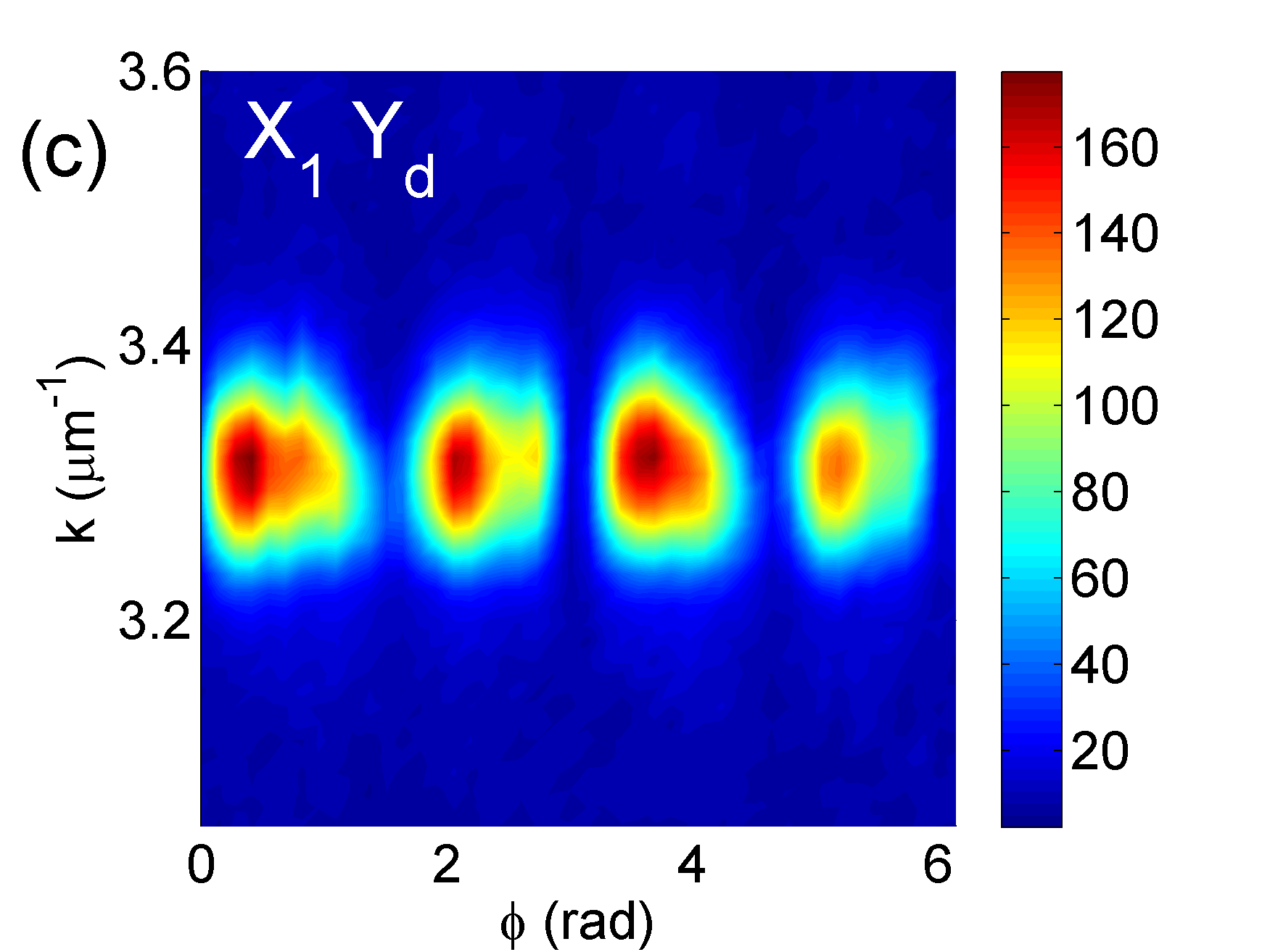}  & \includegraphics[height=3cm]{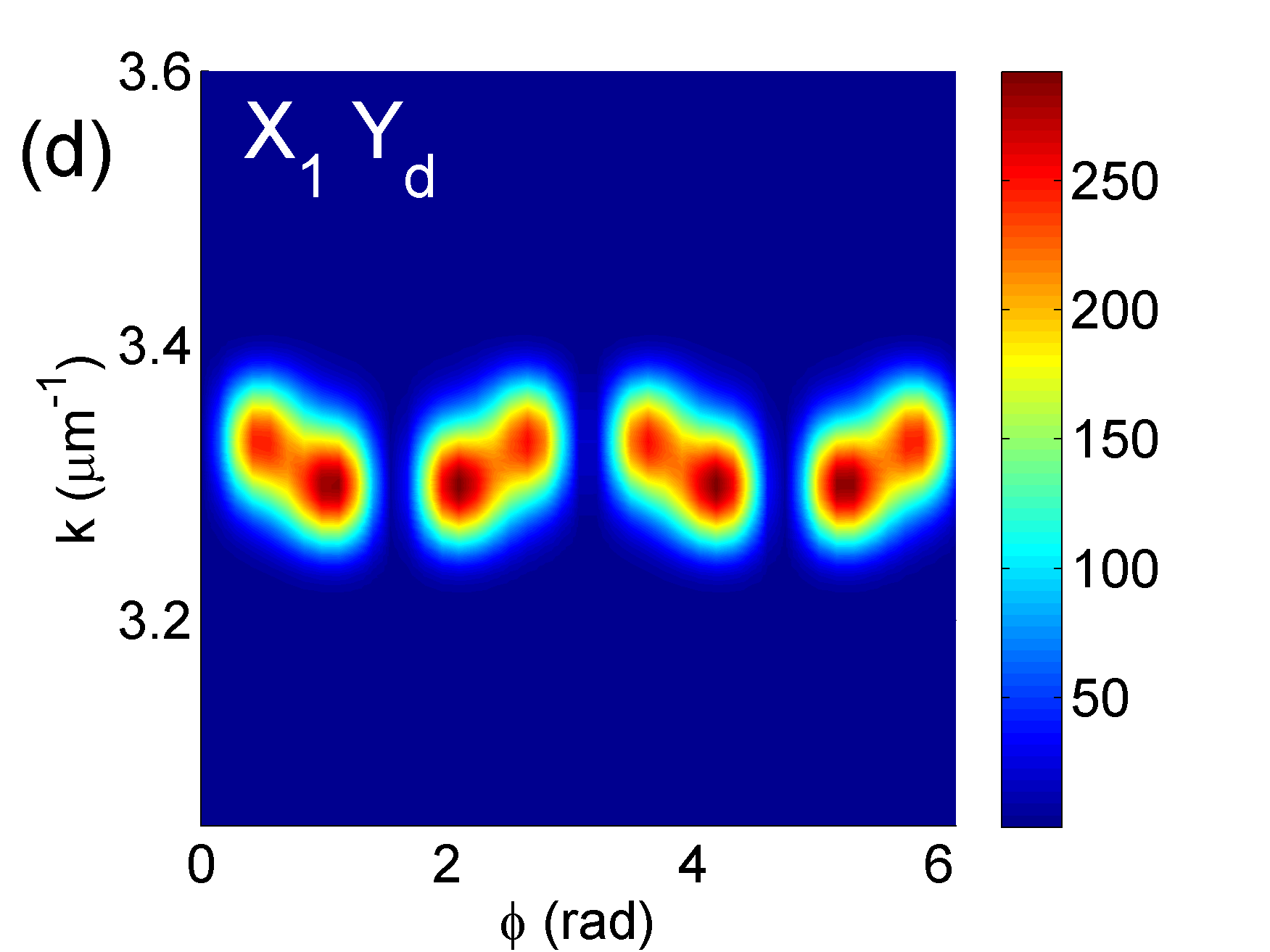}\\
\includegraphics[height=3cm]{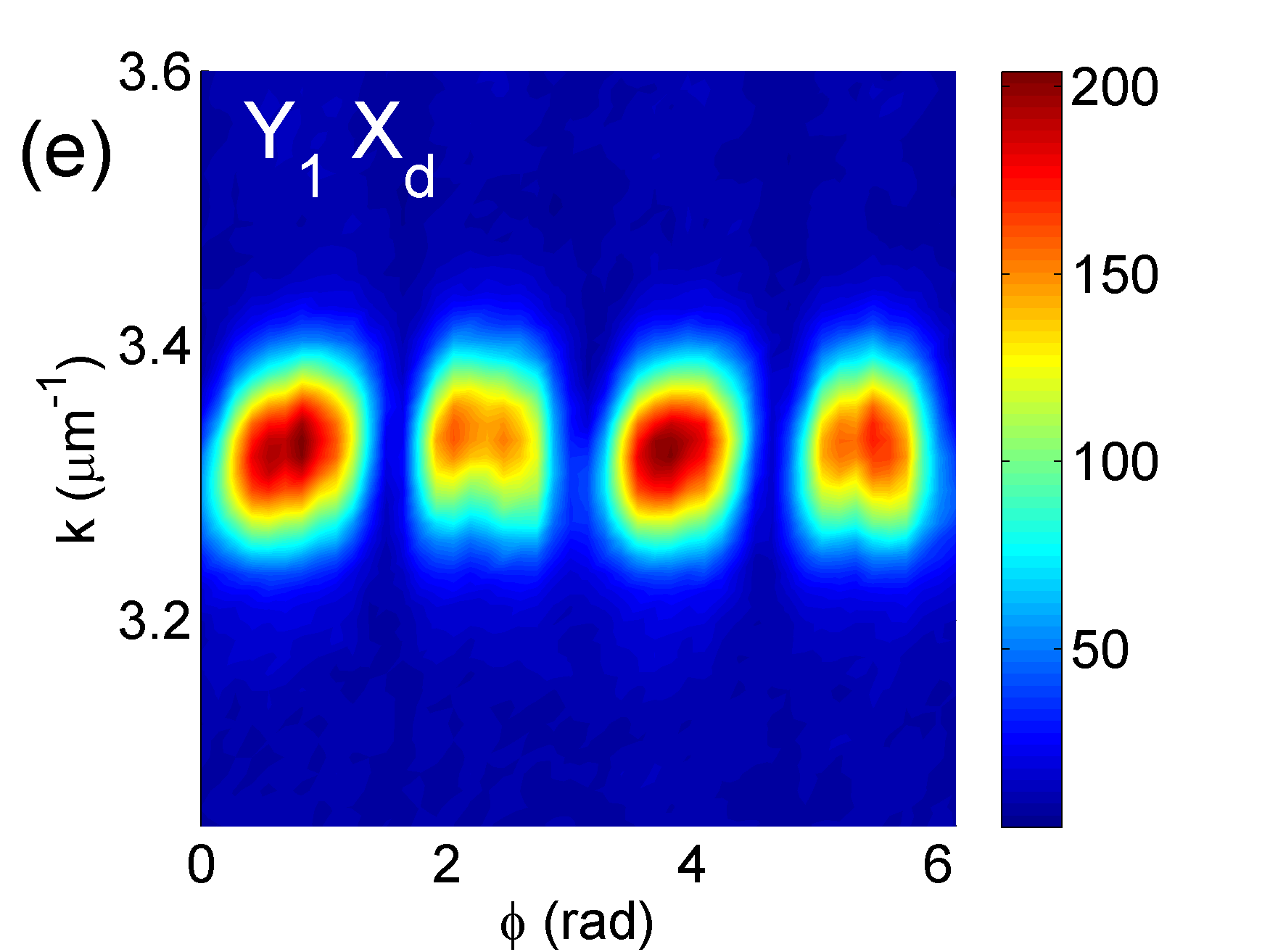}  & \includegraphics[height=3cm]{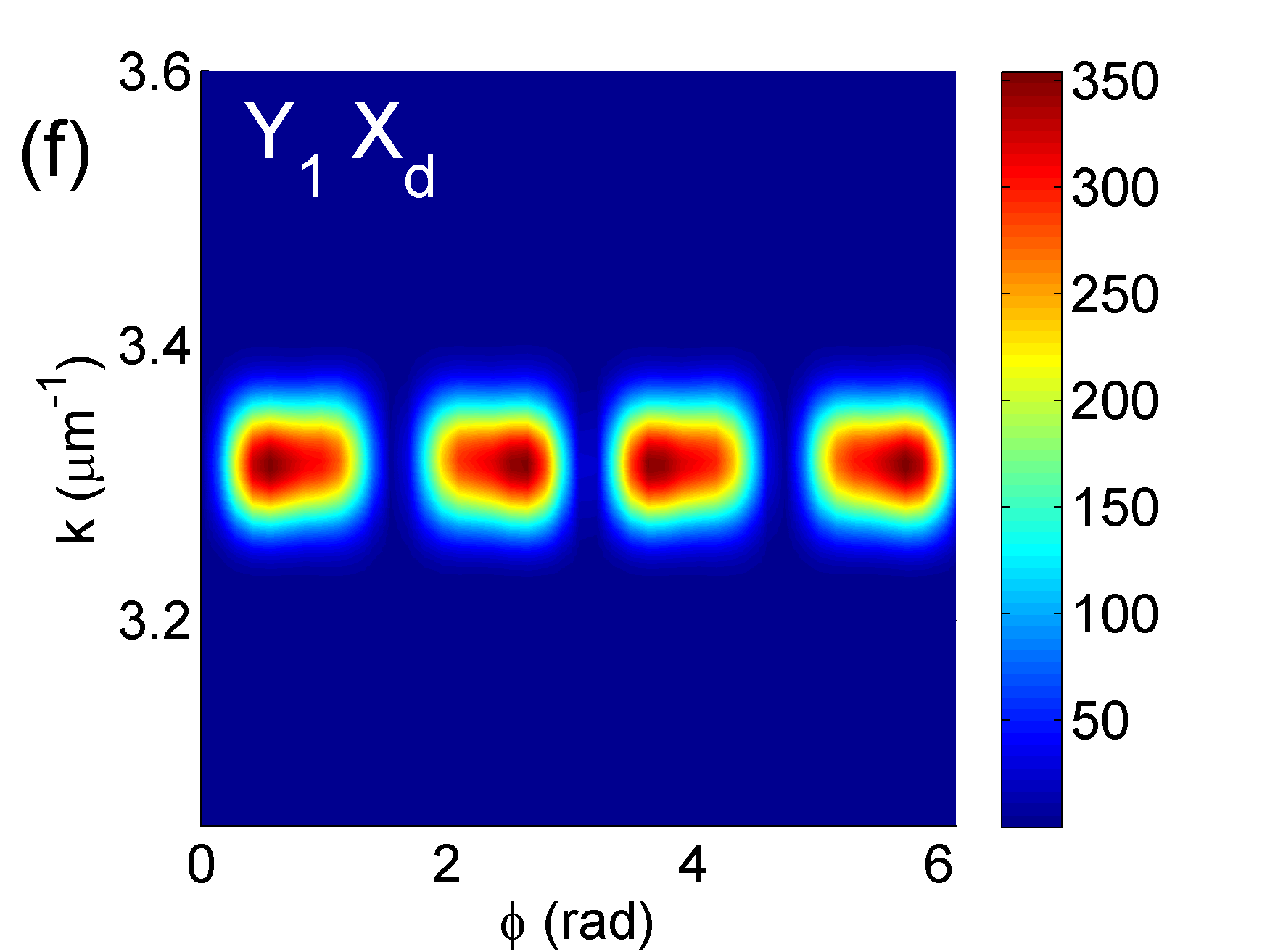}\\
\includegraphics[height=3cm]{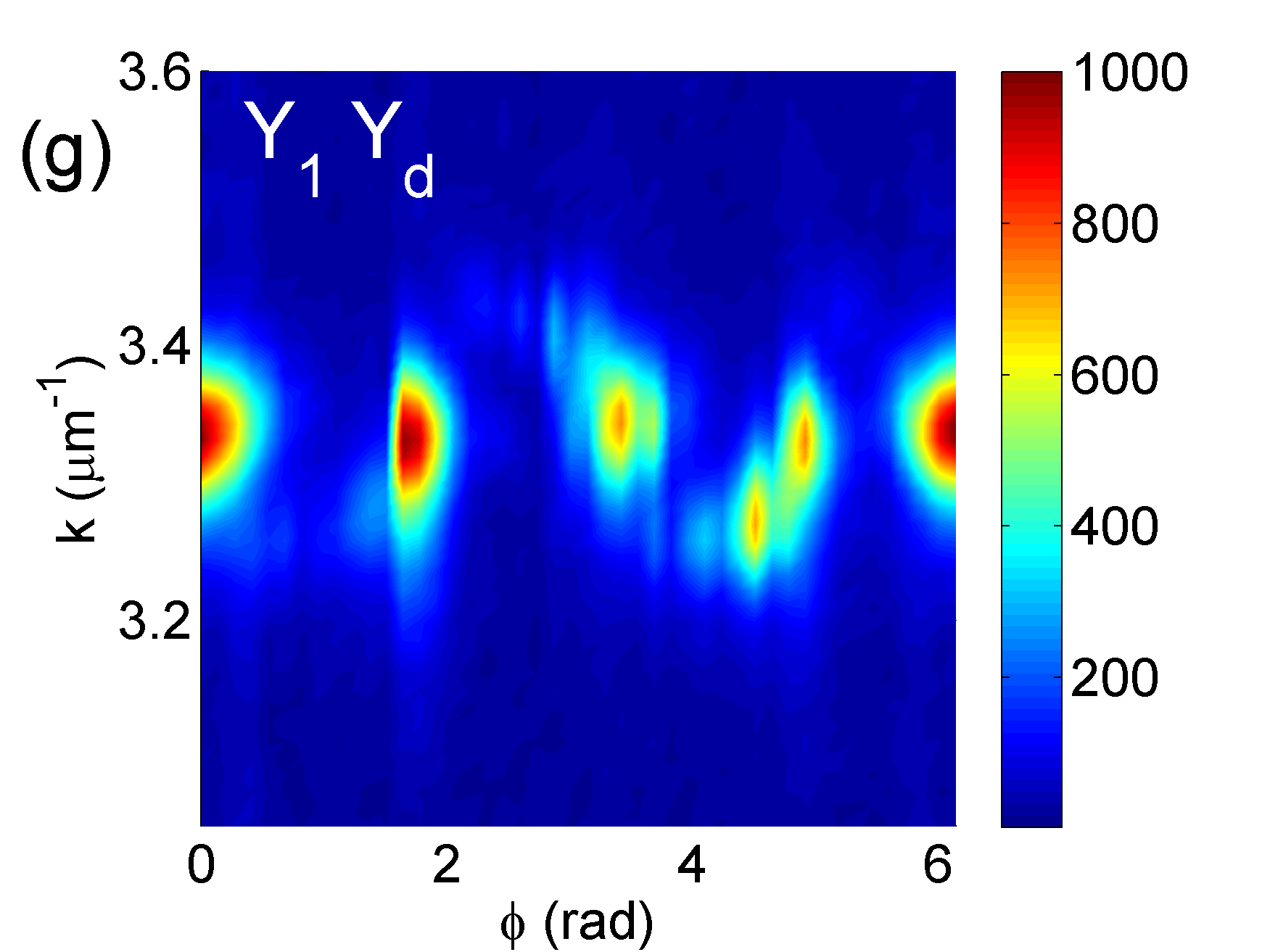}  & \includegraphics[height=3cm]{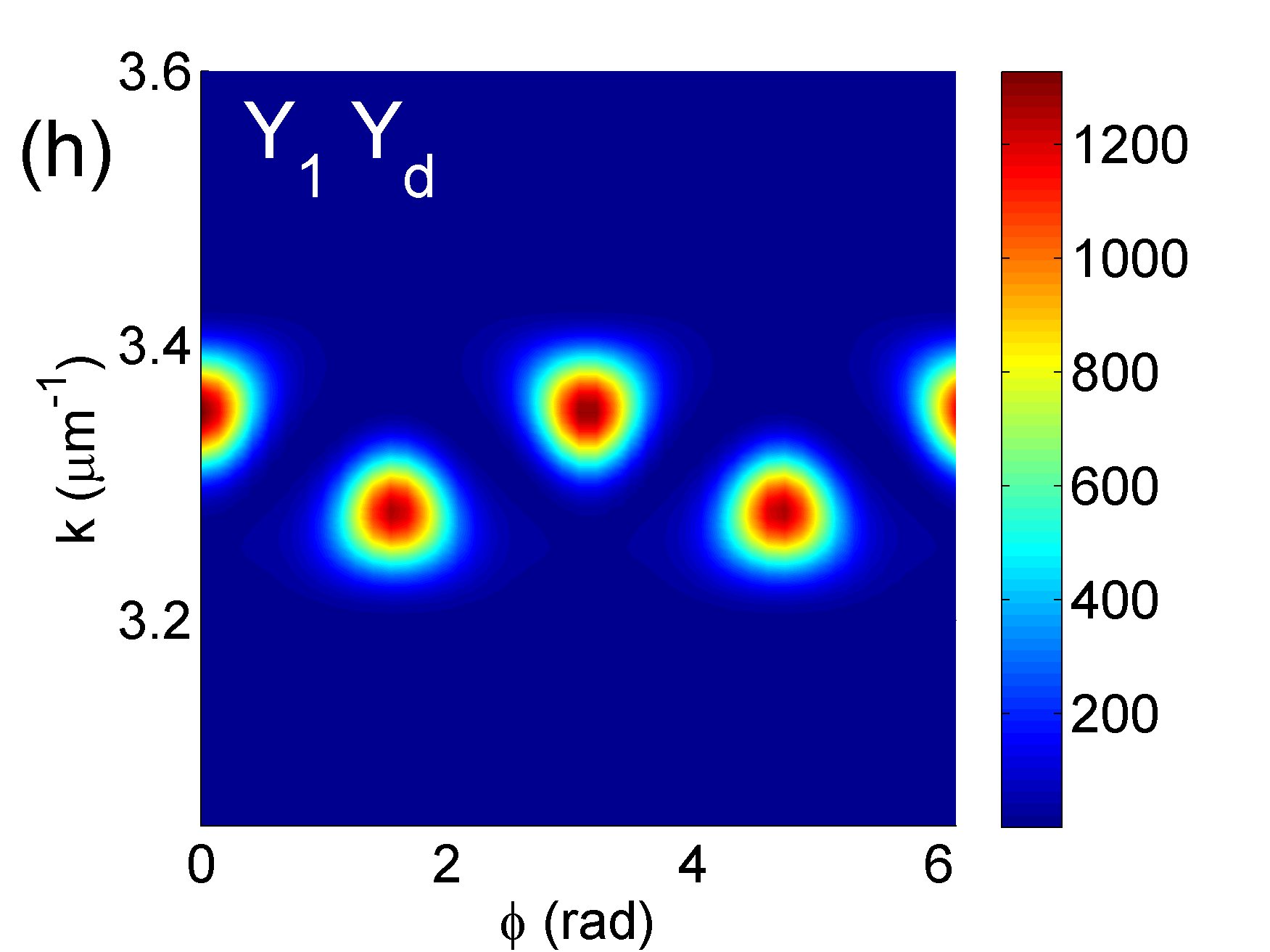}
\end{tabular}
\captionof{figure}{Measured (left column) and computed (right column) first-order four-wave mixing intensity in arbitrary units. Results are shown in the two-dimensional transverse momentum space parametrized by the magnitude of the momentum $k$ and angle $\phi$ which denotes the polar angle between the plane defined by the incident probe and the pump polarization plane (see Fig.~\ref{fig:pump_probe_scheme}a). Each experimental picture is composed of 44 measurements corresponding to vertical stripes on the picture. The polarization configuration in each row is indicated by the symbols for probe (first symbol) and detection (second symbol). In the first two rows, the probe is co-linearly polarized to the pump. In the bottom two rows the probe is cross-linearly polarized to the pump. The filter used in the detection is co-linearly polarized to the pump in rows 1 and 3 and cross-linearly polarized to the pump in rows 2 and 4.}
\label{fig:single_probe_results}
\end{figure}

The experimental and theoretical results are shown in Fig.~\ref{fig:single_probe_results}. The intensity of the FOFWM signal taken for a fixed probe in-plane momentum $\vert \bold{k}_\mathrm{probe}\vert = 3.3 \,\mathrm{\mu m}^{-1}$  is shown as a function of the in-plane momentum $k$ and for each $\phi$ between $0$ and $2\pi$.

The results are shown for four different polarization configurations: (i) probe excitation and FOFWM detection both co-polarized to the pump ($\mathrm{X_1}\mathrm{X_d}$), (ii) a co-polarized probe and a cross-polarized FOFWM-detection to the pump ($\mathrm{X_1}\mathrm{Y_d}$), (iii) and (iv) a cross-polarized probe and co- or cross-polarized FOFWM-detection to the pump, respectively ($\mathrm{Y_1}\mathrm{X_d}$ and $\mathrm{Y_1}\mathrm{Y_d}$). In all four polarization configurations, we find the absence of an azimuthal symmetry of the detected FOFWM signal. Instead, only a two-fold rotational symmetry is found. Furthermore, the detected FOFWM signal does not map the polarization state of the incoming probe beam. Only if the probe incidence is perpendicular or parallel to the pump polarization plane, the probe matches the TE or TM eigenmode, respectively. For each of the polarization configurations, the FOFWM signal exhibits distinct features: In the $\mathrm{X_1}\mathrm{X_d}$ and $\mathrm{Y_1}\mathrm{Y_d}$ excitation/detection-configuration, the FOFWM is strongest if the plane defined by the incident probe is parallel or perpendicular to the pump polarization plane at angles $\phi=0$, $\phi=\frac{\pi}{2}$, $\phi=\pi$, and $\phi=\frac{3\pi}{2}$. For $\mathrm{X_1}\mathrm{X_d}$, the radius of the FOFWM signal is alternating between $k_{\mathrm{FWM}} = 3.25$ $\mu$m$^{-1}$ for parallel and $k_{\mathrm{FWM}} = 3.35$ $\mu$m$^{-1}$ for perpendicular excitation. For $\mathrm{Y_1}\mathrm{Y_d}$, the role of parallel and perpendicular excitation is interchanged. The radius reaches a minimum (maximum) for perpendicular (parallel) excitation. In contrast, in the $\mathrm{X_1}\mathrm{Y_d}$ and $\mathrm{Y_1}\mathrm{X_d}$ configurations the signal vanishes for an excitation of the TE or TM-eigenmode at $\phi=0$, $\phi=\frac{\pi}{2}$, $\phi=\pi$, and $\phi=\frac{3\pi}{2}$, but reaches its maximum for $\phi =\frac{\pi}{4},  \frac{3 \pi}{4}, \frac{5 \pi}{4}$ and  $\frac{7 \pi}{4}$. Here, the variation in radius with varying $\phi$ is much less pronounced, however, a double-peak structure is observed near each signal maximum. Overall, the FOFWM signal is most intense for the $\mathrm{Y_1}\mathrm{Y_d}$-configuration, followed by the $\mathrm{X_1}\mathrm{X_d}$-configuration, and weakest for $\mathrm{X_1}\mathrm{Y_d}$ and $\mathrm{Y_1}\mathrm{X_d}$ configurations. For all these features, qualitative agreement is found between experiment and theory.

However, some artefacts for $\mathrm{Y_1}\mathrm{Y_d}$ configuration are visible in the experimental data (Fig.~\ref{fig:single_probe_results} g). Due to long acquisition times (since each vertical stripe is based on a separate measurement), slight variations in system parameters can occur,  possibly raising the system over the OPO threshold.

In the next section of the present paper we develop a detailed understanding and analyze the features visible in Fig.~\ref{fig:single_probe_results}.
%
%
%
%

\section{Theoretical description}

As mentioned in the introduction, the angular and polarization dependence of the FOFWM signal reported in Fig.~\ref{fig:single_probe_results} is a consequence of the interplay between the TE-TM splitting and spatially anisotropic polariton amplification resulting from the spin-dependent exciton-exciton interaction for a linearly polarized pump.
The former can be understood in the linear optical regime, the latter is a nonlinear optical effect. Both are discussed in detail below and together they consistently explain our observations.
In this section we will work only with a pump with a fixed horizontal polarization, no probe is inserted in the calculations. Therefore, in this section $\phi$ can be simply assimilated to the azimuthal angle on the elastic circle.

\subsection{The linear optical regime} \label{sec:sec:linear-optical-regime}

\begin{figure}
\centering{\includegraphics[height=3.3cm]{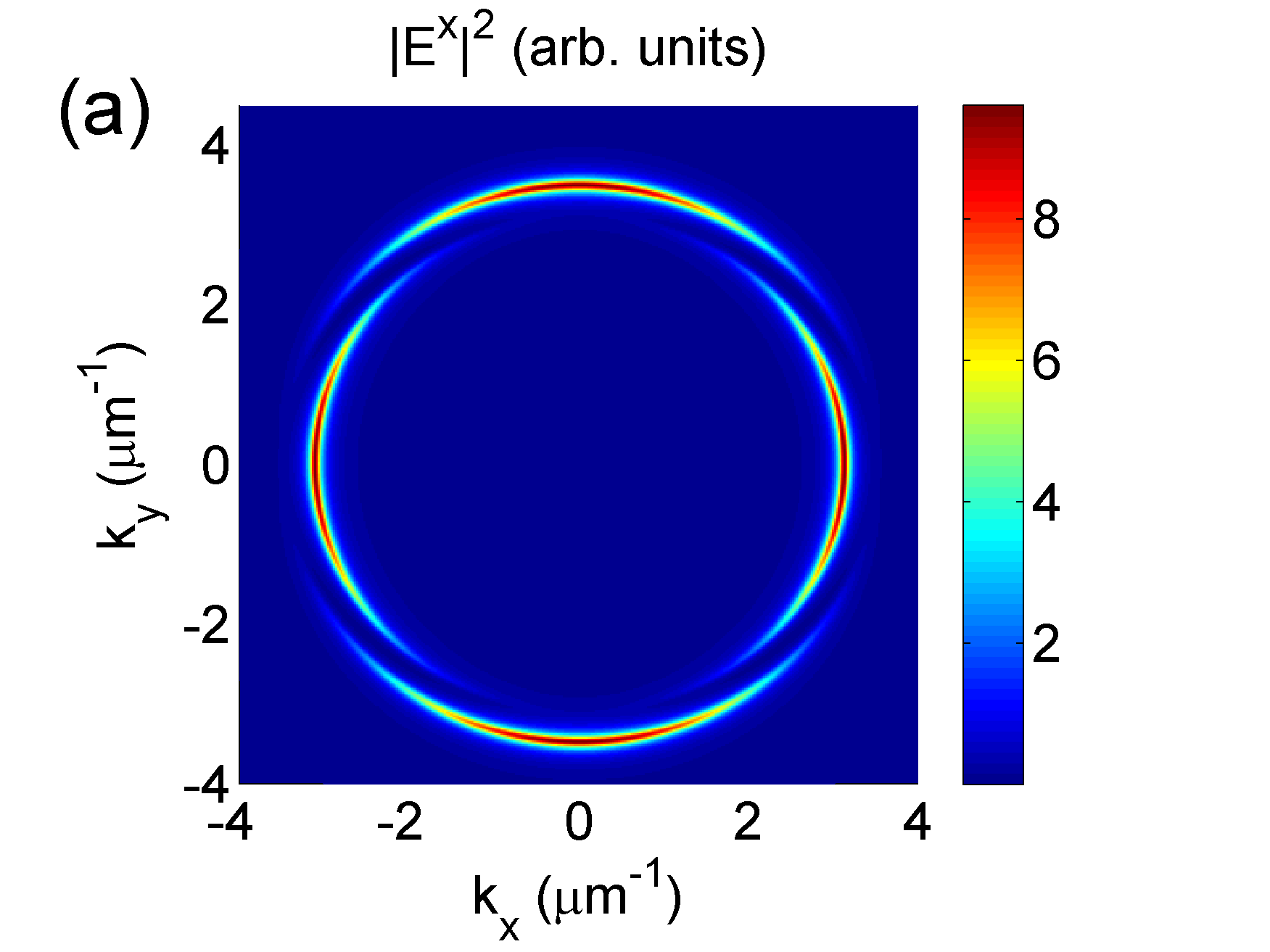}\includegraphics[height=3.3cm]{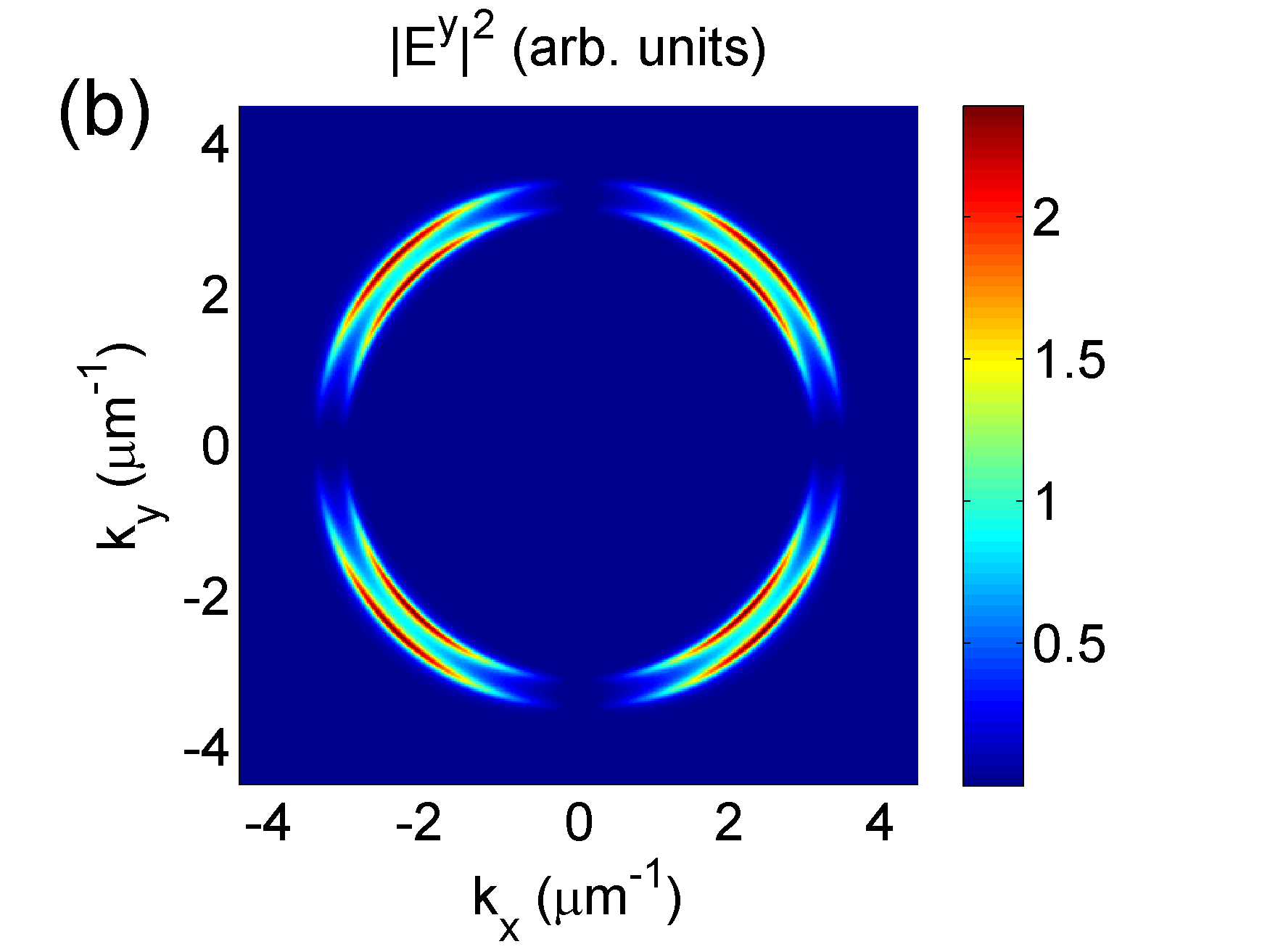}\\
\includegraphics[height=3.3cm]{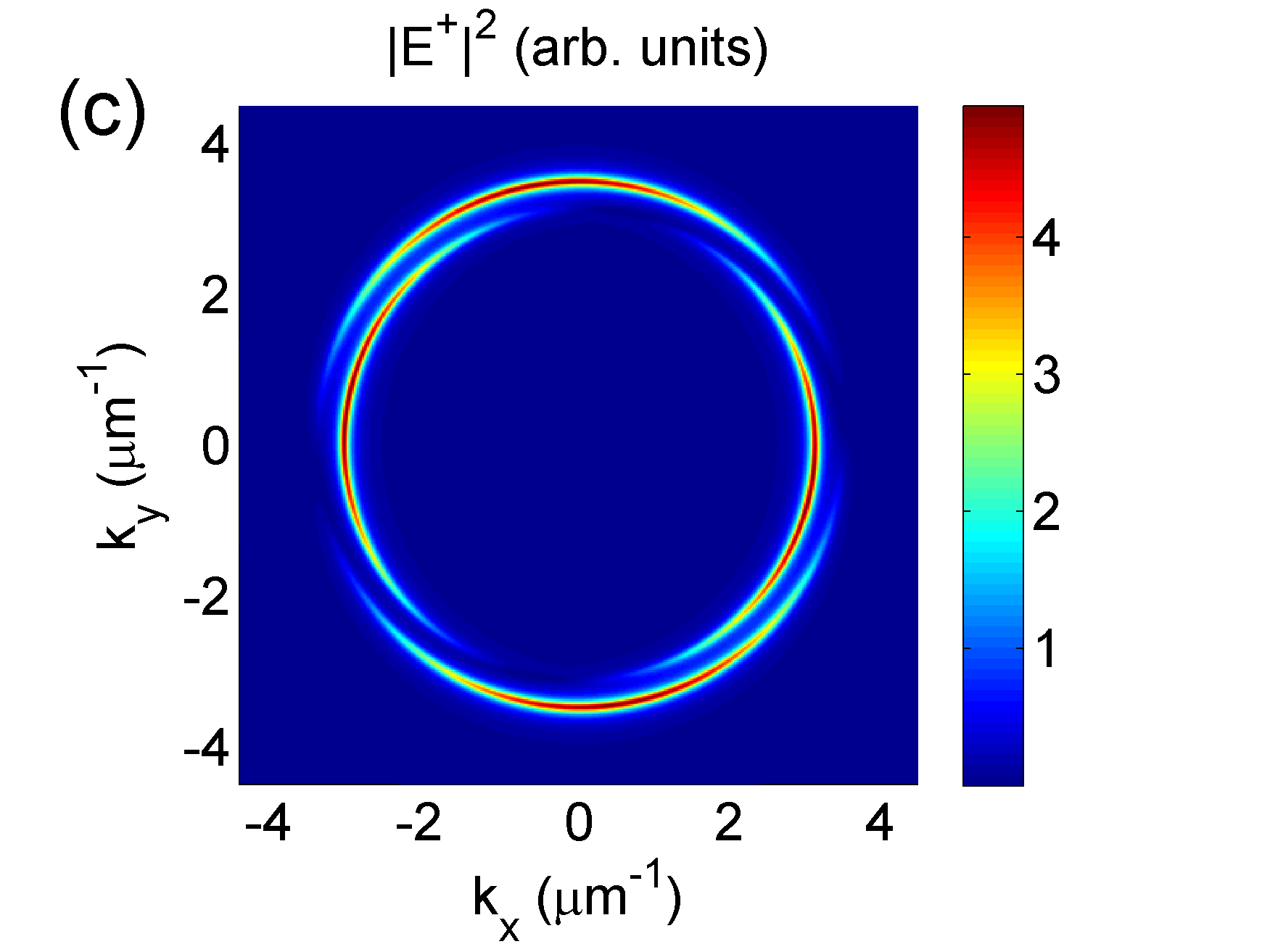}\includegraphics[height=3.3cm]{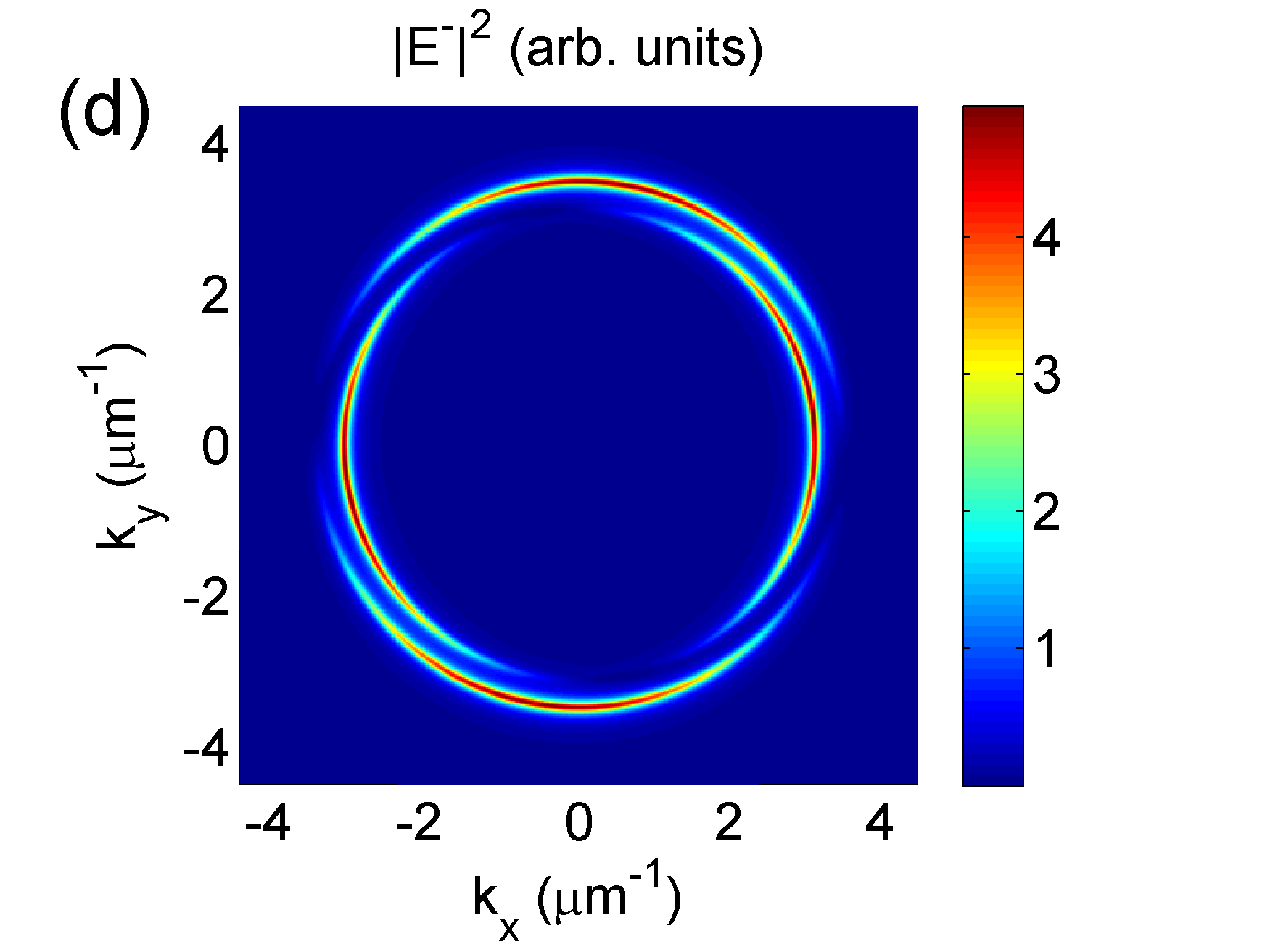}}
\caption{System response in the linear optical regime. Shown is the photon density inside the cavity for excitation with a weak cw pump spread over the whole $k$-space. The pump is linearly polarized with horizontal polarization. Clearly visible is a double-ring structure at $|\mathbf{k}|\approx3.3\,\mathrm{ \mu m^{-1}}$ on which resonant excitation of polaritons on LPB$_1$ occurs. Shown are the (a) co-linearly (X) polarized, (b) cross-linearly (Y) polarized), (c) left-circularly polarized, and (d) right-circularly polarized components of the photonic field. For the linearly polarized pump the TE-TM splitting in the cavity modes breaks the azimuthal symmetry and yields a change of the polarization state. The polarization state of the excitation is only recovered for pump incidence parallel and perpendicular to the polarization axis, respectively.}
\label{fig:lin_properties}
\end{figure}

We start the discussion with a strongly reduced pump intensity. When the excitation intensity is very weak, all nonlinear effects due to exciton-exciton interaction in Eq.~(\ref{eq:GPE_p}) can be neglected and no four-wave mixing signal is observed. For a linearly polarized source entering a cavity with TE-TM splitting under a finite angle, the polarization state of the source light is not strictly conserved upon reflection and transmission. This is the basis for the optical spin Hall effect \cite{Kavokin2005,Leyder2007}.

In order to illustrate this phenomenon based on our system equations, we re-write the photonic part in Eq.~(\ref{eq:GPE_e}) in $k$-space and in the linearly polarized basis. A linearly polarized source is considered. The coupled equations for co- and cross-linear (X and Y, respectively) photonic component then take the following form:
\begin{align}
& \mathrm{i}\hbar \dot{E}_{i,\bold{k}}^{X \atop Y} =  \left(\hbar \omega^c_{k} \pm \cos(2\phi) \Delta_{k}^{TL} - \mathrm{i}\gamma_{c} \right) E_{i,\bold{k}}^{X \atop Y}  -  \Omega_{c}E_{j,\bold{k}}^{X \atop Y} \nonumber  \\ & +  \sin\left(2 \phi \right) \Delta_{k}^\mathrm{TL} E_{i,\bold{k}}^{Y \atop X}  -  \Omega_{x}p_{i,\bold{k}}^{X \atop Y}+E_{\mathrm{pump},i,\bold{k}}\delta_{{X \atop Y},X}\,.
\label{eq:GPE_E_field}
\end{align}
Here, $\hbar \omega^c_k={\hbar \omega^c_0 +} {{\hbar^2 k^2}\over{4}}\left(\frac{1}{m_\mathrm{TM}} + \frac{1}{m_\mathrm{TE}} \right)$ is the cavity dispersion without transverse-longitudinal splitting. $2 \Delta_{k}^\mathrm{TL}$ denotes the $k$-dependent TE-TM splitting. 
As can be seen in Eq.~(\ref{eq:GPE_E_field}), the Y component $E_{i,\bold{k}}^{Y}$ is driven by the X-polarized field component $E_{i,\bold{k}}^{X}$. The strength of this source term depends on the angle $\phi$, which leads to a momentum- and angular-dependent change of the polarization state of the incoming light when entering the cavity.
In the linear optical regime, the excitonic part (Eq.~(\ref{eq:GPE_p})) simplifies to
\begin{eqnarray}
\mathrm{i} \hbar \dot{p}_{i,\bold{k}}^{X \atop Y} & = & \left( \epsilon^x_0 - \mathrm{i} \gamma_x \right) p_{i,\bold{k}}^{X \atop Y} - \Omega_x E_{i,\bold{k}}^{X \atop Y}\,.
\label{eq:GPE_p_field}
\end{eqnarray}
Under stationary cw excitation conditions, all dynamical quantities oscillate with the pump frequency $\omega_\mathrm{pump}$ and the coupled equations, Eqs.~(\ref{eq:GPE_E_field}) and (\ref{eq:GPE_p_field}), can be solved for each $\bold{k}$ independently. Fig.~\ref{fig:lin_properties} shows the calculated photonic field for monochromatic excitation with the same  amplitude $A$ for each $k$-point $E_{\mathrm{pump},i,\bold{k}}=A \cdot \exp(-\mathrm{i} \omega_\mathrm{pump} t)$,  i.e. the pump is spread homogenousely over the whole $k$-space.
For better visibility both losses $\gamma_c = \gamma_x = 0.2 \, \mathrm{meV}$ and TE-TM splitting $m_{\mathrm{TE}} = 1.25 \cdot m_{\mathrm{TM}}$ are artificially increased in Fig.~\ref{fig:lin_properties}. {\color{black} For the visualization, we assume the emission from the $\mathrm{LPB_1}$ polariton branch to be symmetric.} Fig.~\ref{fig:lin_properties}a shows the intensity of the X-polarized cavity field, Fig.~\ref{fig:lin_properties}b the Y-polarized intensity and in Fig.~\ref{fig:lin_properties}c and \ref{fig:lin_properties}d the intensity after projection onto the left- and right-circularly polarized components is shown, respectively.  For $\phi = 0,{{\pi}\over{2}},\pi$ and ${{3 \pi}\over{2}}$, the polariton field has no cross-linear component, and the colinear component reaches its maximum (cf. Figs.~ \ref{fig:lin_properties}a and  \ref{fig:lin_properties}b). In these cases, the polarization state of the incident light matches the transverse and longitudinal eigenmodes. For any other angle, there is a cross-linear contribution with an intensity proportional to $\vert \sin\left(2\phi \right) \vert^2$, i.e., reaching its maxima for $\phi = {{\pi}\over{4}},{{3 \pi}\over{4}},{{5 \pi}\over{4}}$ and ${{7 \pi}\over{4}}$. TE-TM splitting lifts the azimuthal symmetry of the resonance condition on the polariton branch LPB$_1$ and the elastic circle splits into crescents with a fourfold symmetry as evident in Figs.~\ref{fig:lin_properties}a and b. Due to a phase shift between the co- and cross-linear component, the excited polariton field is not linearly polarized anymore, but elliptically: the projection on the circular polarization components is not symmetric, as shown in Figs.~\ref{fig:lin_properties}c and d. The degree of elliptization quantitatively depends on $\phi$, on the strength of TE-TM splitting, and on the dephasing $\gamma_c$ and $\gamma_x$. We find in agreement with previous studies \cite{Kavokin2005, Leyder2007, Langbein2007} that TE-TM splitting alone already breaks the azimuthal symmetry and polarization conservation in the linear optical regime. However, TE-TM splitting only does not fully explain the polarization dependence we find for the first order FWM signals in the previous section, e.g. the relative intensities of each configuration of Fig.~\ref{fig:single_probe_results}.
In the second step of our discussion, we go beyond the strictly linear optical regime and analyze the polarization selective FWM in terms of a linear stability analysis (LSA) of Eqs.~(\ref{eq:GPE_e}) and (\ref{eq:GPE_p}).
%
%
%
%
\subsection{The nonlinear optical regime}
\label{sec:sec:LSA} 

\begin{figure*}
\begin{tabular}[t]{lrr}
\includegraphics[scale=0.4]{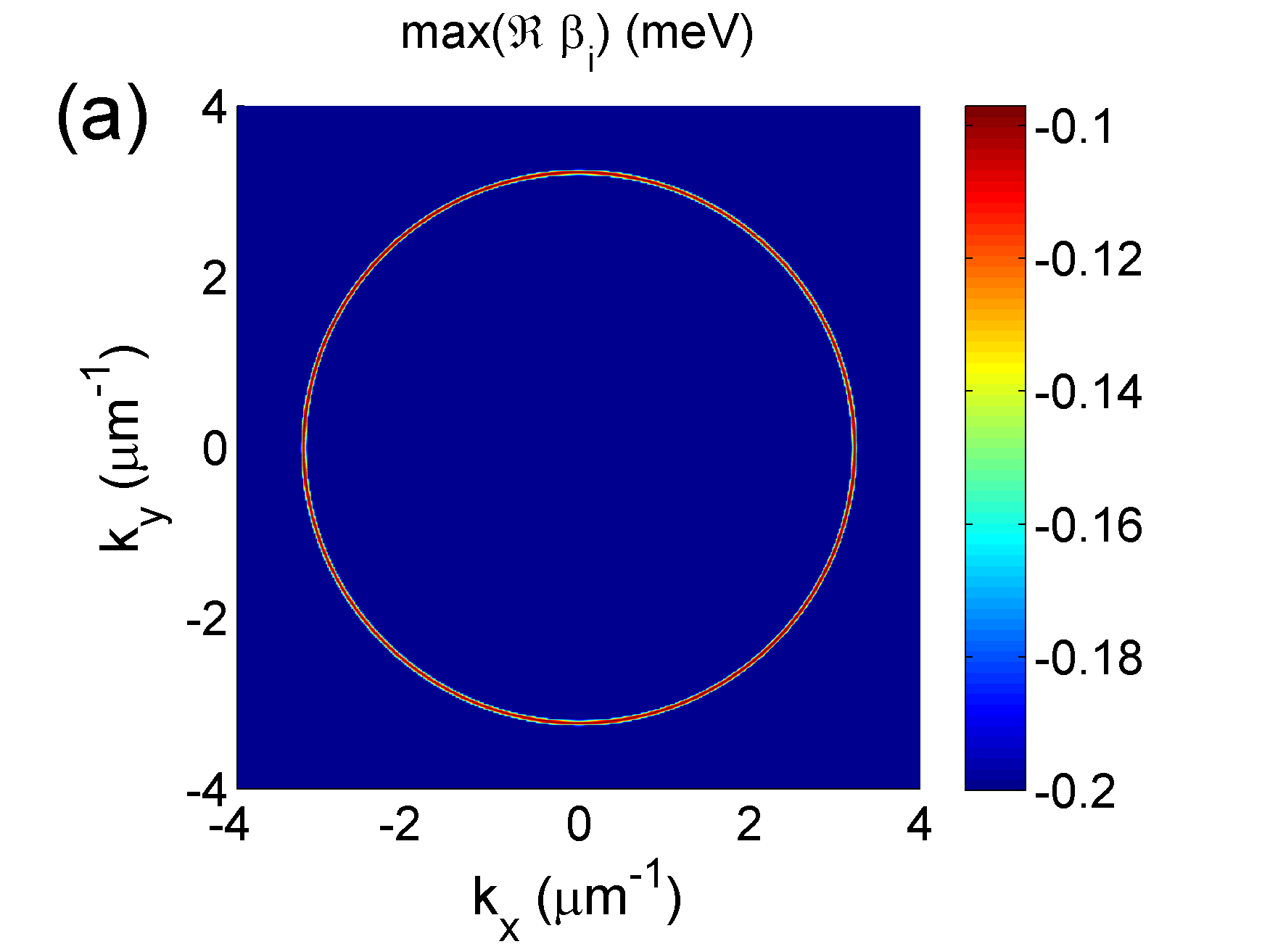} & \includegraphics[scale=0.4]{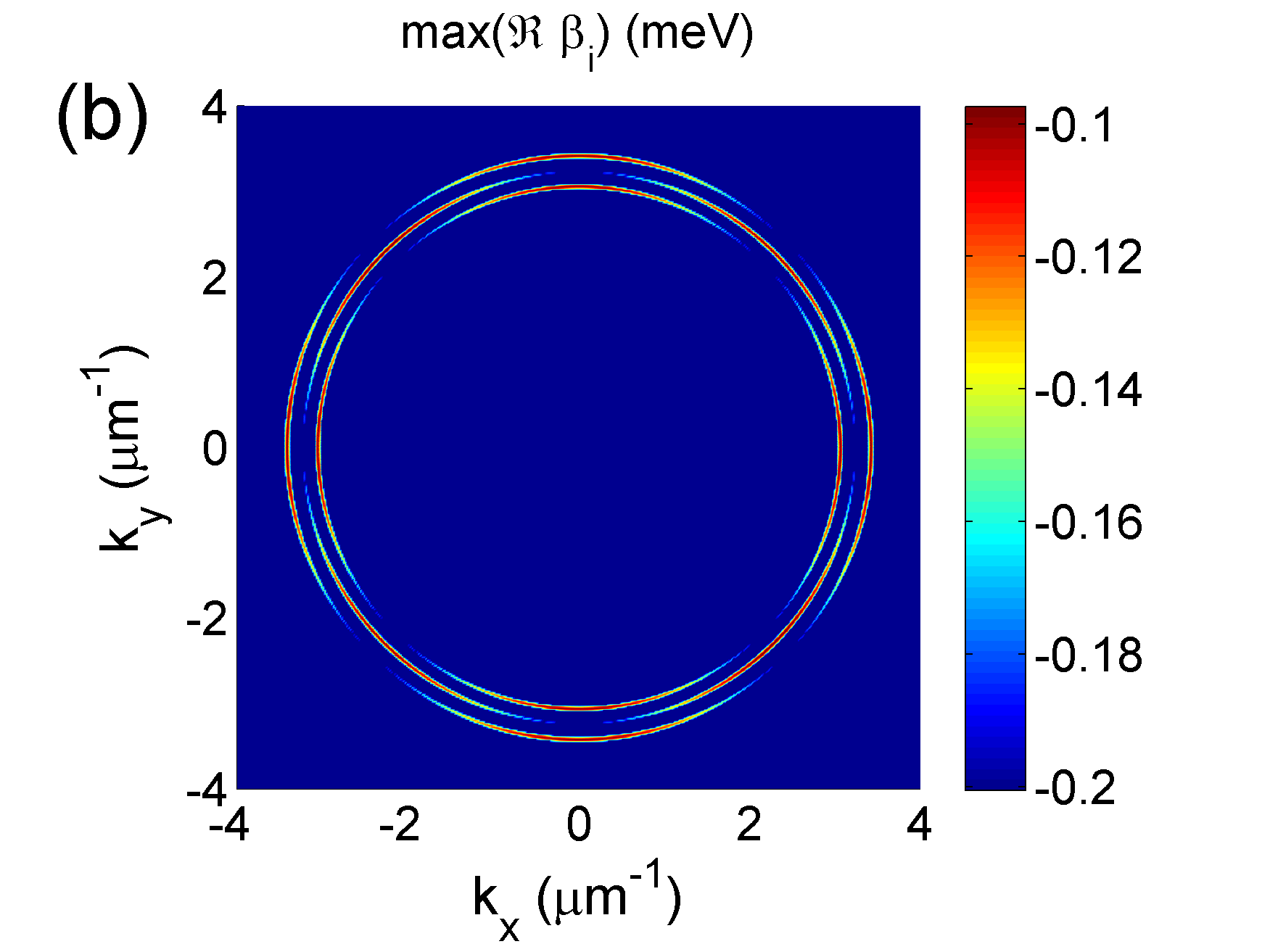} & \includegraphics[scale=0.4]{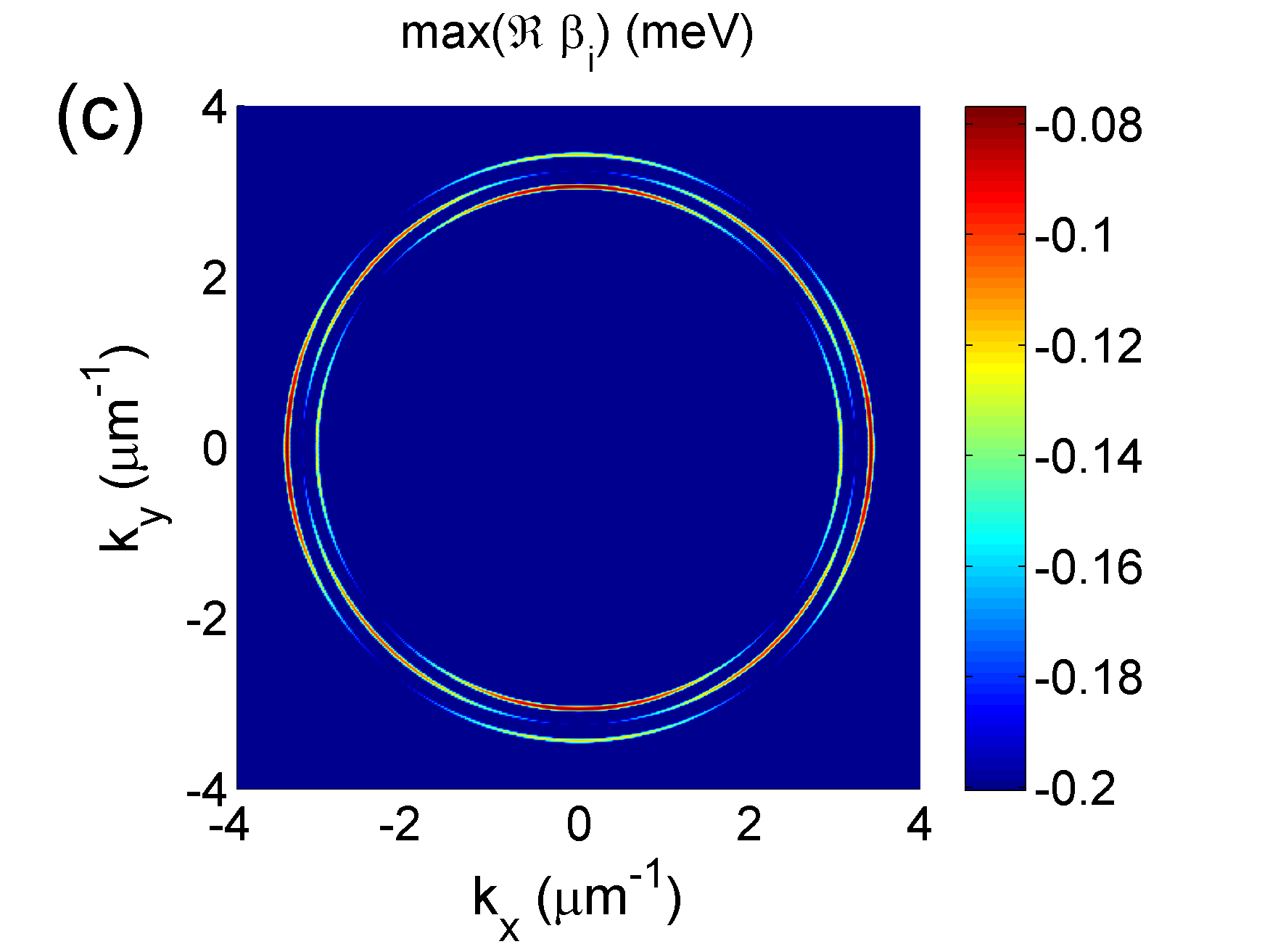}\\
\includegraphics[scale=0.4]{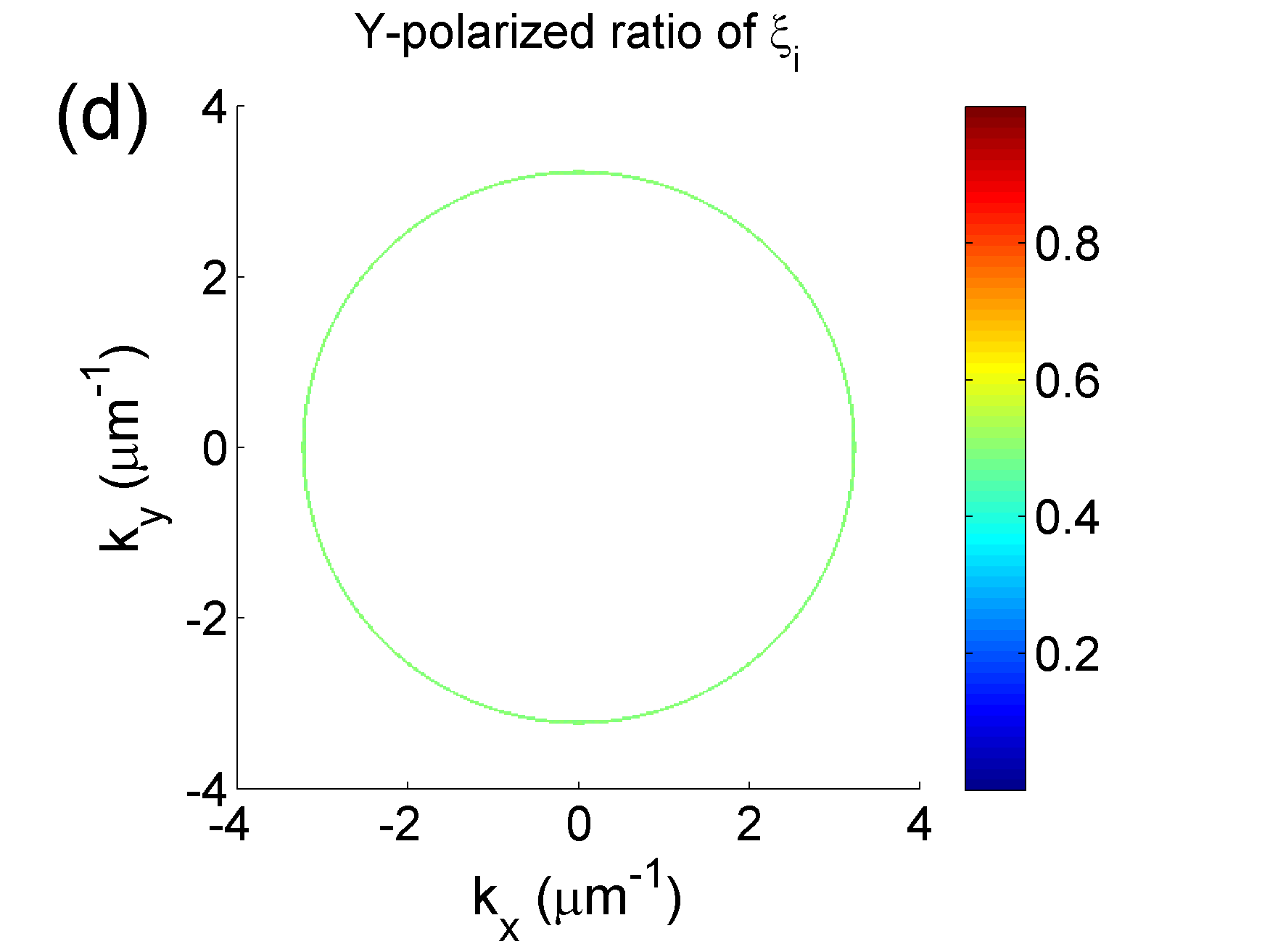} & \includegraphics[scale=0.4]{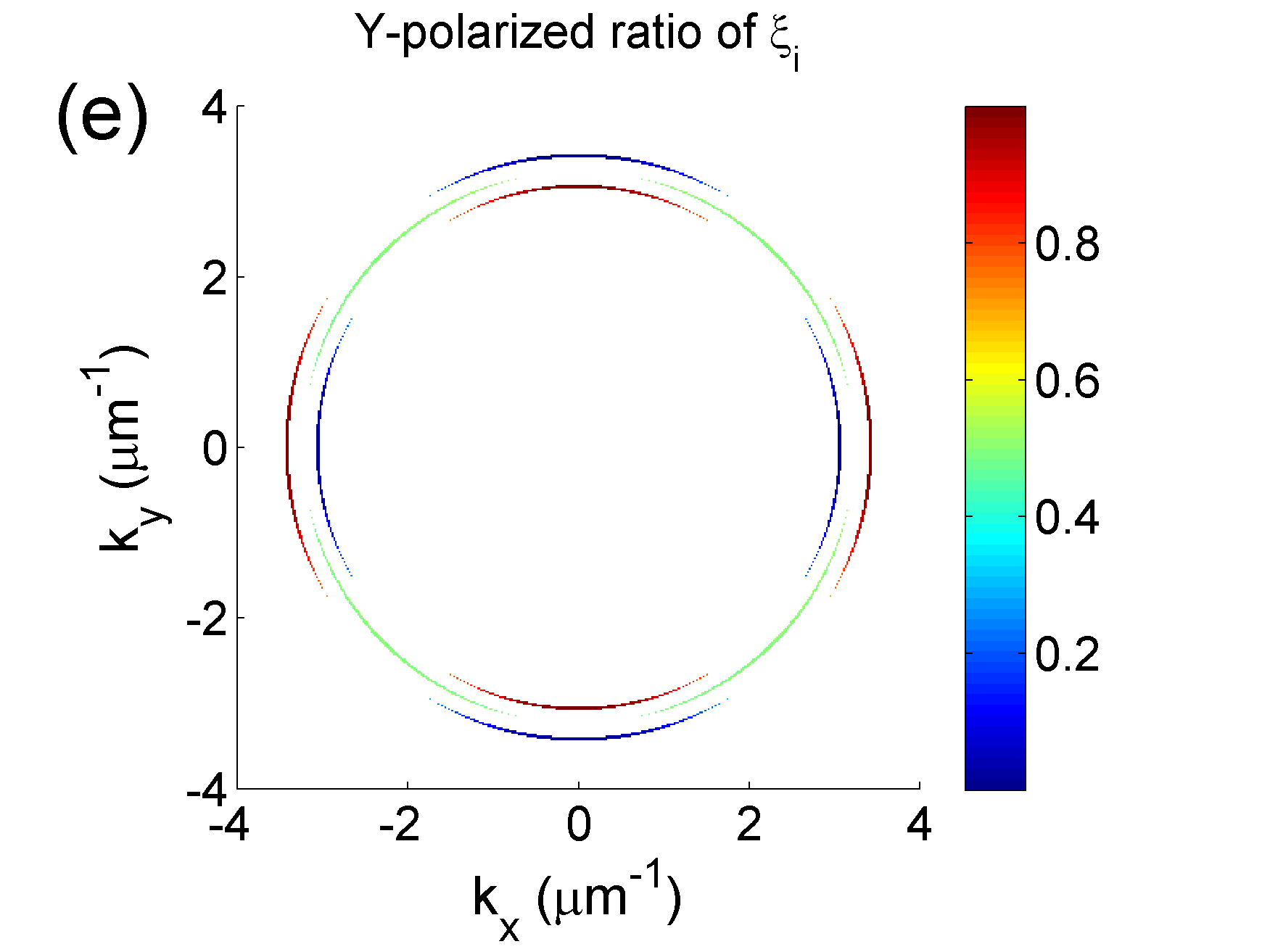} & \includegraphics[scale=0.4]{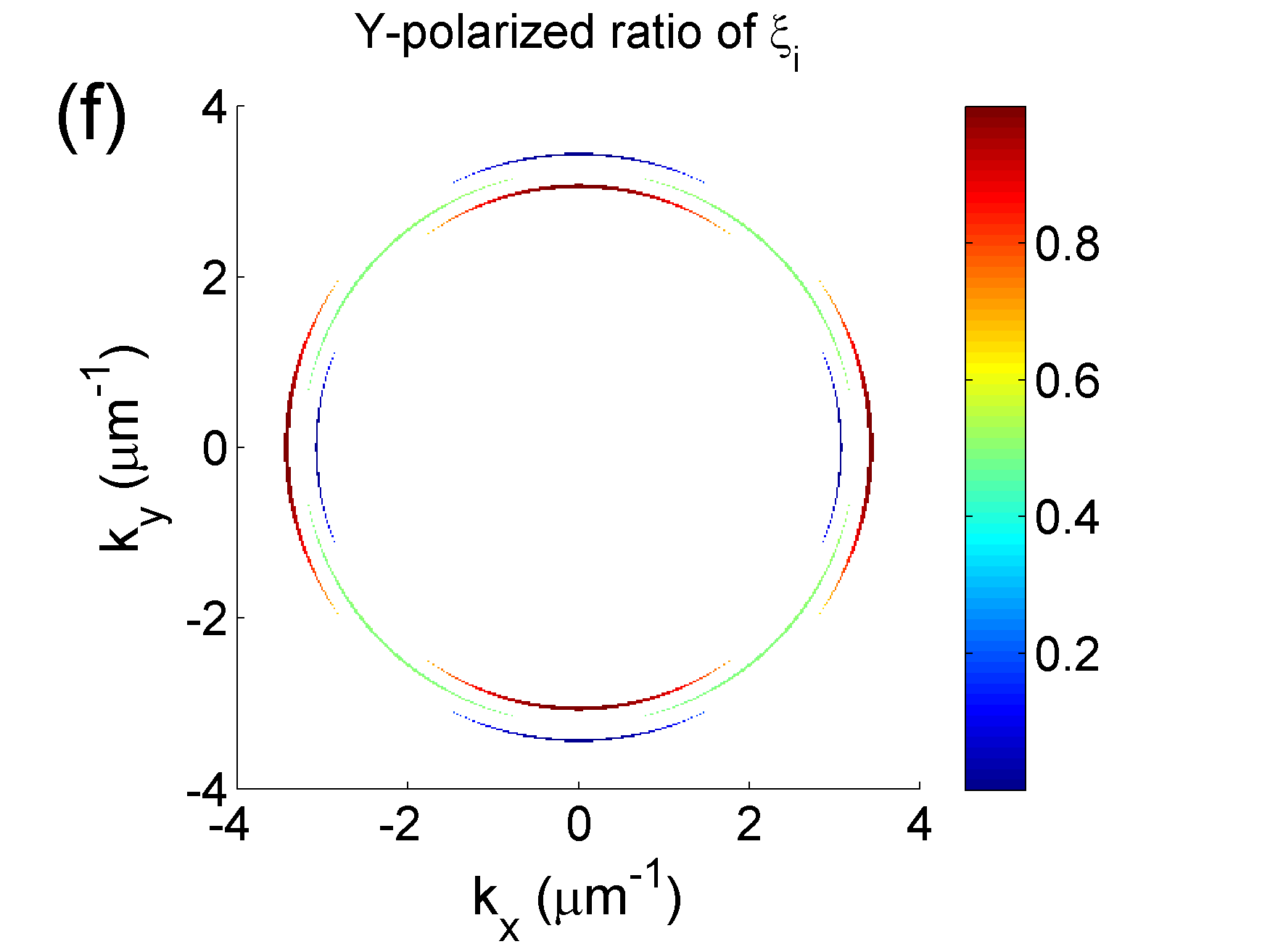}
\end{tabular}
\caption{Results of the linear stability analysis. In the upper row, the real part of the eigenvalue $\beta_i$ with the largest real part is shown. We stay in the regime $\mathfrak{Re}\{\beta_i(\mathbf{k})\}<0$ such that for each $\mathbf{k}$ the slowest decaying mode is shown. The continuous-wave pump is at $\mathbf{k}=0$ and polarized in the horizontal direction. The interaction between cocircularly excitons, $T^{++}$ is kept unchanged. Shown are the following configurations: (a) for zero TE-TM splitting and $T^{+-}=0$, (b) for exaggerated TE-TM splitting with $m_\mathrm{TE}=1.25  \cdot m_\mathrm{TM}$ and $T^{+-}=0$, (c) for exaggerated TE-TM splitting $m_\mathrm{TE}=1.25 \cdot m_\mathrm{TM}$ and $T^{+-}<0$. (d) to (f) show the projection of the corresponding eigenmode $\xi_i$ on the Y-polarization state. For clarity, only the range in the vicinity of the elastic circle is shown where $\mathfrak{Re}\{\beta_{i}\}>-0.2$ meV. Comparing corresponding plots in upper and lower row, clearly visible is that where $\mathfrak{Re}\{\beta_i(\mathbf{k})\}$ takes its largest values, the signal is predominantly Y-polarized.}
\label{fig:LSA}
\end{figure*} 

To study the  polarization dependent FOFWM in detail, again we consider a linearly polarized,  monochromatic excitation. In contrast to sec.~3 A, the pump is spread homogenousely over the whole cavity plane, i.e. $E_{\mathrm{pump},i}(\bold{r},t) = A\cdot\exp(-\mathrm{i} \omega_\mathrm{pump} t)$.
In the linear optical regime discussed above, any polariton field decays over time with $\gamma_p$, determined by the dephasing of the photonic and excitonic components, $\gamma_c$ and $\gamma_x$, respectively. 
In the presence of a strong pump beam, however, off-axis signals on the elastic circle or a finite-amplitude probe may also be amplified by pairwise resonant scattering of pump polaritons into the probe $\bold{k}$ and first-order four-wave mixing direction $-\bold{k}$.

Due to this stimulated amplification, the decay of off-axis polaritons is effectively reduced \cite{Ballarini2009Observation}. When the amplification outweighs the polariton loss, even exponential growth of off-axis signals can be observed, rendering the spatially homogenous polariton field unstable \cite{Ardizzone2013,Luk2013, Saito2013, Egorov2014}. In the present work we stay below this instability threshold such that four-wave mixing can be systematically probed.
\begin{figure}
\hspace*{-2cm}\includegraphics[scale=0.5, trim= 0cm 6.5cm 0cm 5.0cm]{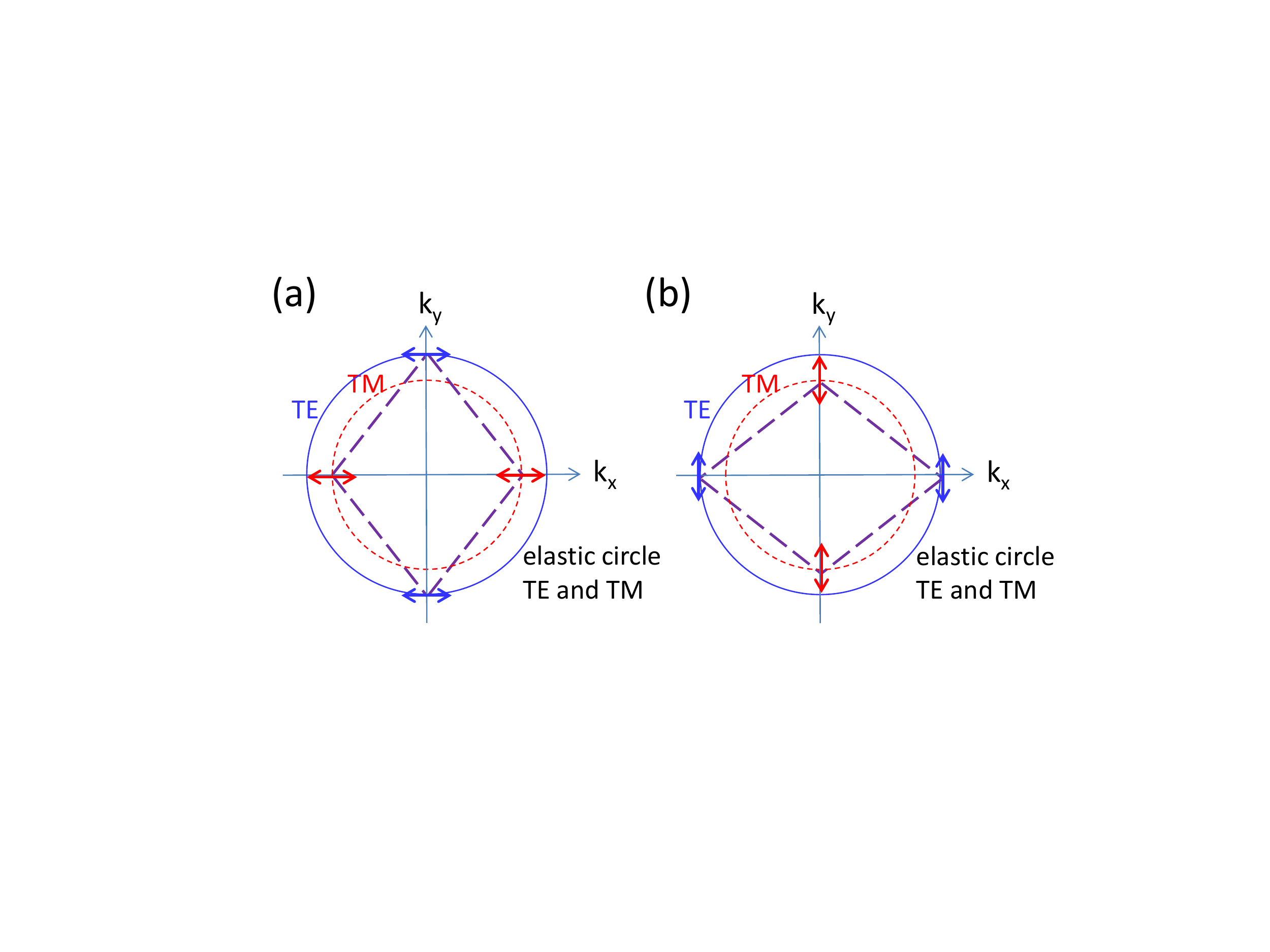} 
\captionof{figure}{Sketch of the two-dimensional transverse momentum space (k-space) plane and linear polarization states indicated by double arrows. The pump is horizontally polarized. The states that are purely X or purely Y polarized (shown in a and b, respectively) are highlighted by connecting lines. Exciton-exciton interaction favors scattering onto Y-polarized states. }
\label{fig:sketch-mostlyX-mostlyY}
\end{figure} 

In this regime, the effective decay of off-axis polaritons can be analytically obtained based on a linear stability analysis (LSA) \cite{Schumacher2007a,Schumacher2007c}. In this approach, we approximate the polariton field as a superposition of a pump at $\bold{k}_\mathrm{pump}=0$ (comprising the cavity field $E_{i,0}^{\pm}$ and the excitonic polarization $p_{i,0}^{\pm}$), an off-axis field at $\bold{k}$ ($E_{i,\bold{k}}^{\pm}$, $p_{i,\bold{k}}^{\pm}$) and its corresponding conjugate first-order four-wave mixing field at $-\bold{k}$ ($E_{i,-\bold{k}}^{\pm}$, $p_{i,-\bold{k}}^{\pm}$), all oscillating at the pump frequency: $E_{i}^{\pm}(\bold{r},t)=\left(E_{i,0}^{\pm}+E_{i,\bold{k}}^{\pm}(t)\mathrm{e}^{\mathrm{i} \bold{k}\bold{r}}+E_{i,-\bold{k}}^{\pm}(t)\mathrm{e}^{-\mathrm{i} \bold{k} \bold{r}}\right)\mathrm{e}^{-\mathrm{i}\omega_\mathrm{pump} t}$
and $p_{i}^{\pm}(\bold{r},t)=\left(p_{i,0}^{\pm}+p_{i,\bold{k}}^{\pm}(t)\mathrm{e}^{\mathrm{i} \bold{k} \bold{r}}+p_{i,-\bold{k}}^{\pm}(t)\mathrm{e}^{-\mathrm{i}\bold{k}\bold{r}}\right)\mathrm{e}^{-\mathrm{i}\omega_\mathrm{pump} t}$.
To analyze the stability of a stationary solution of Eqs.~(\ref{eq:GPE_e}) and (\ref{eq:GPE_p}) for the case when only the spatially homogeneous pump field is fixed, only $E_{i,\bold{k}}^{\pm}$, $E_{i,-\bold{k}}^{\pm}$, $p_{i,\bold{k}}^{\pm}$ and $p_{i,-\bold{k}}^{\pm}$ remain as dynamical degrees of freedom. If we insert this ansatz into the equation of motion, Eqs.~(\ref{eq:GPE_e}) and (\ref{eq:GPE_p}) for each in-plane momentum $\bold{k}$, the coupled dynamic of 16 field components (the photonic and excitonic field at $\bold{k}$ and $\bold{-k}$ in each cavity and in each polarization state) results in:
\begin{equation}
\frac{\partial}{\partial t}\left(\begin{array}{c}
\mathbb{E}_{1}\\
\mathbb{P}_{1}\\
\mathbb{E}_{2}\\
\mathbb{P}_{2}\end{array}\right)=-\frac{\mathrm{i}}{\hbar}\left(\begin{array}{cccc}
\mathbb{M}^1_{EE} & \mathbb{M}^1_{Ep} & \mathbb{M}_{12} & 0\\
\mathbb{M}^1_{pE} & \mathbb{M}^1_{pp} & 0 & 0\\
\mathbb{M}_{21} & 0 & \mathbb{M}^2_{EE} & \mathbb{M}^2_{Ep}\\
0 & 0 & \mathbb{M}^2_{pE} & \mathbb{M}^2_{pp}\end{array}\right)\left(\begin{array}{c}
\mathbb{E}_{1}\\
\mathbb{P}_{1}\\
\mathbb{E}_{2}\\
\mathbb{P}_{2}\end{array}\right).
\label{eq:LSA}
\end{equation}
Since we are interested in the linear stability of the pump-induced stationary solution, off-axis field amplitudes in order higher than one are neglected in the derivation. For clarity, the 16 components are organized in quatuples $\left(\mathbb{E}_{1},\mathbb{P}_{1},\mathbb{E}_{2},\mathbb{P}_{2}\right)^{T}$ with $\mathbb{E}_{i}=\left(E_{i,\bold{k}}^{+},E_{i,\bold{k}}^{-},E_{i,-\bold{k}}^{+*},E_{i,-\bold{k}}^{-*}\right)^{T}$
and $\mathbb{P}_{i}=\left(p_{i,\bold{k}}^{+},p_{i,\bold{k}}^{-},p_{i,-\bold{k}}^{+*},p_{i,-\bold{k}}^{-*}\right)^{T}$. Details of the 16x16 matrix $\mathbb{M}$ are given in
Appendix A. The time dependent solution $\Xi \equiv \left(\mathbb{E}_{1},\mathbb{P}_{1},\mathbb{E}_{2},\mathbb{P}_{2}\right)$ of Eq.~(\ref{eq:LSA})
can be decomposed into the eigenstates $\xi_{i}$ of $\mathbb{M}$,
i.e. $\Xi=\sum_{i=1\ldots16}\lambda_{i}\xi_{i}$, with $\mathbb{M}\xi_{i}=\beta_{i}\xi_{i}$, where $\beta_{i}$ are the eigenvalues of $\mathbb{M}$. The dynamics described by
Eq.~(\ref{eq:LSA}) can then be expressed through the dynamics in each eigenmode from an initial condition at $t=0$, $\lambda_{i}(t)=\exp(\beta_{i}t)\lambda_{i}(t=0)$. The real part of the eigenvalues $\beta_i$, $\mathfrak{Re}(\beta_{i})$, governs exponential growth ($\beta_{i}>0$) or decay ($\beta_{i}<0$). A finite imaginary part, $\mathfrak{Im}(\beta_{i})$, leads to frequency shifts from the pump frequency.

For a given initial condition of Eq.~(\ref{eq:LSA}), different polarization components can be amplified differently based on their overlap with the eigenvectors $\xi_i$ of $\mathbb{M}$ and the corresponding eigenvalues $\beta_i$ for each mode. In general, this will lead to a change in polarization state over time through pump-induced amplification as considered here. For sufficiently strong pumping, the long-time behavior obtained from Eq.~(\ref{eq:LSA}) is dominated by the eigenmode with the slowest decay, i.e., largest $\mathfrak{Re}(\beta_{i})$ (if the initial condition has a finite field contribution in this eigenmode). In Fig.~\ref{fig:LSA} we give a detailed analysis of this mode with the slowest decay rate for fixed exciton density induced by a linearly  horizontally-polarized pump. {\color{black} For clarity, only the relevant eigenmodes with frequencies ($|\mathfrak{Im}(\hbar \beta_{i})|< 1$ meV) are included in the visualization.} The upper row shows $\max_{i}\left(\mathfrak{Re}(\beta_{i})\right)$ in the $k$-domain. The bottom row shows the projection of the corresponding eigenstate $\xi_i$ onto the linear polarization state that is orthogonal (Y polarized) to the pump polarization state. To disentangle the role that TE-TM splitting and excitonic interactions play for the results shown in Fig.~\ref{fig:LSA}, below we give a step by step discussion of results for three different scenarios. The results presented in Fig.~\ref{fig:LSA} are obtained by numerical diagonalization of the 16x16 matrix $\mathbb{M}$ in Eq.~(\ref{eq:LSA}). A more intuitive and analytical picture of the general symmetry considerations underlying these results can be given based on a simplified model as detailed in Appendix B.

In Fig.~\ref{fig:LSA}a, TE-TM splitting is switched off and the interaction between excitons with opposite circular polarization, $T^{+-}$, is zero. Since the resonance and phase matching conditions for off-axis scattering of pump-induced polaritons are best fulfilled there, the elastic circle marks the range with the strongest amplification (or lowest effective loss). Everywhere else in $k$-space, this process is off-resonant and therefore amplification is inefficient such that the effective decay of polaritons coincides with the intrinsic loss included in the model. Furthermore, without TE-TM splitting and for $T^{+-}=0$, the amplification is independent of the probe polarization state: X and Y component of the LSA-eigenstates are balanced (note that only the Y-component is shown in Fig.~\ref{fig:LSA}d).
In Fig.~\ref{fig:LSA}b, TE-TM splitting is included but still with $T^{+-}=0$. For clarity, the TE-TM splitting is exaggerated here. Clearly visible in Fig.~\ref{fig:LSA}b TE-TM splitting lifts the azimuthal symmetry into a fourfold symmetry. The resonant circle visible in Fig.~\ref{fig:LSA}a now splits into twelve crescents at three different radii. Besides the broken symmetry in the lifetime of polariton modes, the distribution of polarization components of $\xi_i$ changes remarkably. Each of the four pairs of crescents parallel and perpendicular to the polarization axis is mostly X- or Y-polarized, respectively (see Fig.~\ref{fig:sketch-mostlyX-mostlyY} for a schematic view of purely X and purely Y polarized states). The remaining four crescents (on the diagonals) are in an elliptical polarization state, where the projection onto X- and Y-component is balanced (note that in Fig.~\ref{fig:LSA}e only the Y component is shown).
%
%
In Fig.~\ref{fig:LSA}c and~f TE-TM splitting is the same as in Fig.~\ref{fig:LSA}b and e but in addition a finite coupling of excitons with opposite circular polarization is included, with $T^{+-}=-T^{++}/3<0$ for negative detuning from the exciton resonance \cite{Schumacher2007a}. The polarization state of $\xi_i$ (Fig.~\ref{fig:LSA}f) remains the same as in Fig.~\ref{fig:LSA}e, but the eigenvalues $\mathfrak{Re}(\beta_{i})$ take different maximum values. The fourfold symmetry visible in Fig.~\ref{fig:LSA}b is reduced into a twofold symmetry in Fig.~\ref{fig:LSA}c. Also, parallel and orthogonal to the pump polarization, in each pair of crescents the two crescents take very different values. This  corresponds to a preferred scattering of pump polaritons and signal growth in those off-axis modes polarized perpendicular to the pump.

\subsection{Discussion}
\label{sec:sec:Discussion} 
Coming back to the experimental and theoretical results presented in section~\ref{sec:single-probe-results}, these can now be fully understood considering the interplay of both linear and nonlinear properties discussed in this section. Using an X-polarized probe beam and detecting the X-polarized signal ($\mathrm{X_1 X_d}$ configuration), FOFWM can be stimulated most efficiently in the TE- and TM-eigenmode for probe incidence in X- and Y direction, respectively. This leads to the alternating radius of the resonance peaks in the $\mathrm{X_1 X_d}$-configuration (cf.~Fig.~\ref{fig:lin_properties}a). In analogy, there is an alternating radius for the peaks in the $\mathrm{Y_1 Y_d}$ configuration, but with interchanged roles of the TE- and TM-modes. Since a Y-polarized probe (that is cross-polarized to the pump) induces a stronger off-axis scattering of pump polaritons (cf. Fig.~\ref{fig:LSA}c), we find a more intense FOFWM signal in $\mathrm{Y_1 Y_d}$ configuration than in $\mathrm{X_1 X_d}$.
For angles where the probe polarization matches the TE or TM-eigenmode, also the polarization state of the FOFWM matches the polarization of the incident probe. Hence, there is no signal in $\mathrm{X_1 Y_d}$ and $\mathrm{Y_1 X_d}$ at angles $\phi=0$, $\phi=\frac{\pi}{2}$, $\phi=\pi$, and $\phi=\frac{3\pi}{2}$. Instead, the resonance peaks for these configurations are centered near $\phi = \frac{\pi}{4}, \frac{3 \pi}{4}, \frac{5 \pi}{4}$ and $\frac{7\pi}{4}$, where the TE-TM splitting leads to the strongest rotation of the polarization plane (cf.~Fig.~\ref{fig:lin_properties}b). Including the interplay with the polarization selective scattering of pump polaritons, a splitting of these peaks into a two-peak fine structure is found for both $\mathrm{X_1 Y_d}$ and $\mathrm{Y_1 X_d}$ configurations in experiment and theory.

\begin{figure}
\includegraphics[height=6cm]{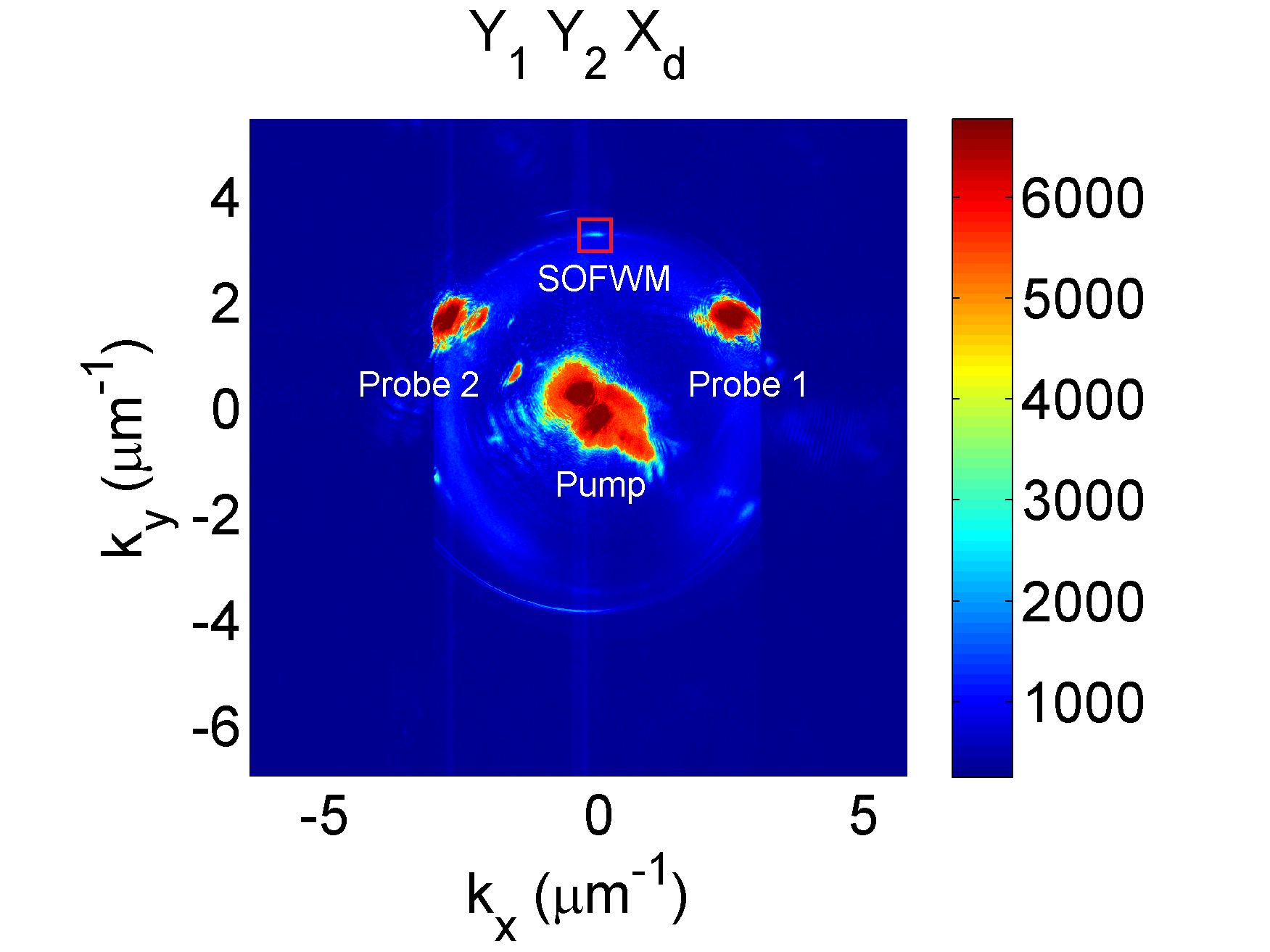}
\caption{\textcolor{black}{Measured second-order four-wave mixing with a pump and two probe beams for two Y-polarized probe beams in X-polarized detection. The SOFWM in hexagonal geometry is marked by a surrounding red rectangle. The field of view is cut on the sides by the spectrometer input slit.}}
\label{fig:double_probeexp_scheme}
\end{figure}

\section{Second order four-wave mixing} 
\label{sec:sec:double_probe} 
In this section we go beyond the analysis of nonlinear processes that are predominantly based on first-order FWM, i.e. the pairwise scattering of two pump polaritons into off-axis modes \cite{Buck2004}.

The system at hand is very well suited to also study four-wave mixing that is of higher order in the off-axis signal. To this end, in addition to the pump and probe beam, we introduce a second probe beam to systematically induce a second-order four-wave mixing signal. Both probes again have the same frequency as the pump and the angle of incidence is chosen such that they are resonant with the lowest polariton branch LPB$_1$. Their polar angle of incidence differs by 120$^{\circ}$, which allows phase-matched and resonant second-order four-wave mixing on the elastic circle and arises in hexagonal geometry with respect to both probe beams as shown in Fig. \ref{fig:pump_probe_scheme}.
At high pump intensities this process crucially contributes to the formation and stabilization of hexagonal patterns \cite{Ardizzone2013,dawes-etal.10,Saito2013}.

For a scalar polariton field, SOFWM has been discussed in detail, e.g., in \cite{Luk2013}. In a spinor field, however, apart from momentum and energy conservation, also spin conservation plays an important role for the four-wave mixing to be efficient. Spin selection rules of the second order four-wave mixing can directly be studied by varying the polarization states of the incoming probe beams. In order to conduct polarization selective experiments, we again work below the instability threshold.

\begin{figure}				
\includegraphics[height=3cm]{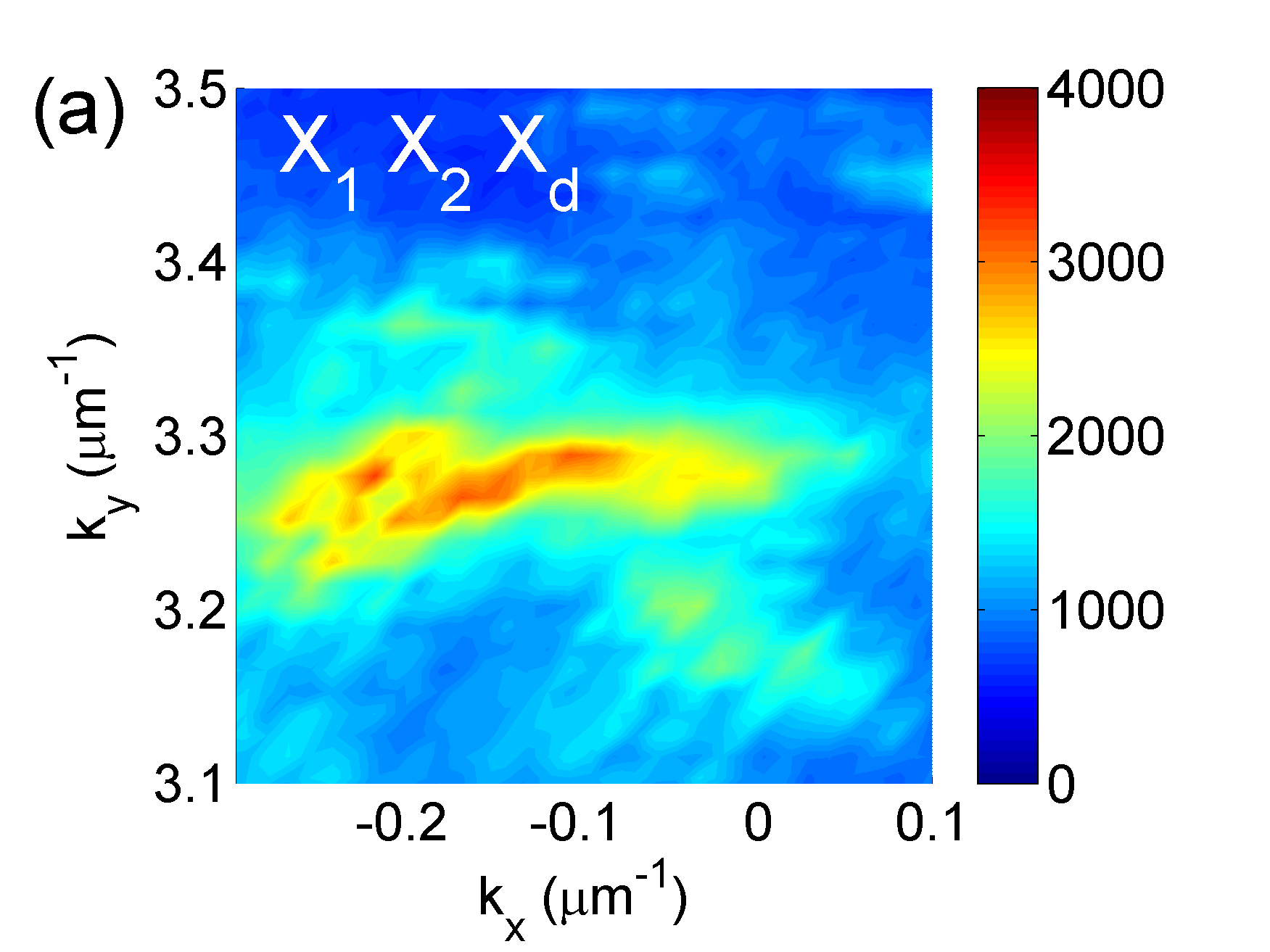} \includegraphics[height=3cm]{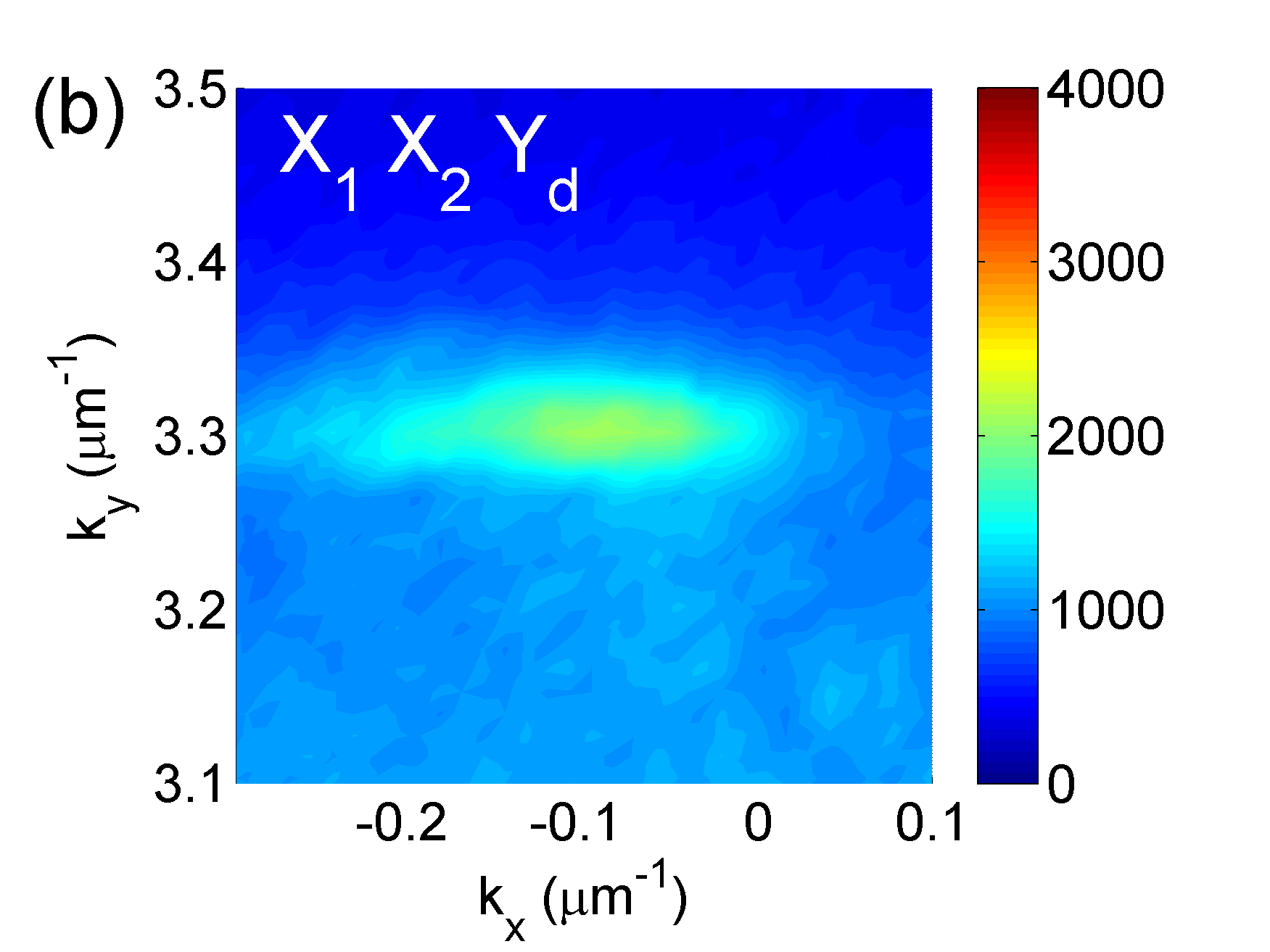} 
\includegraphics[height=3cm]{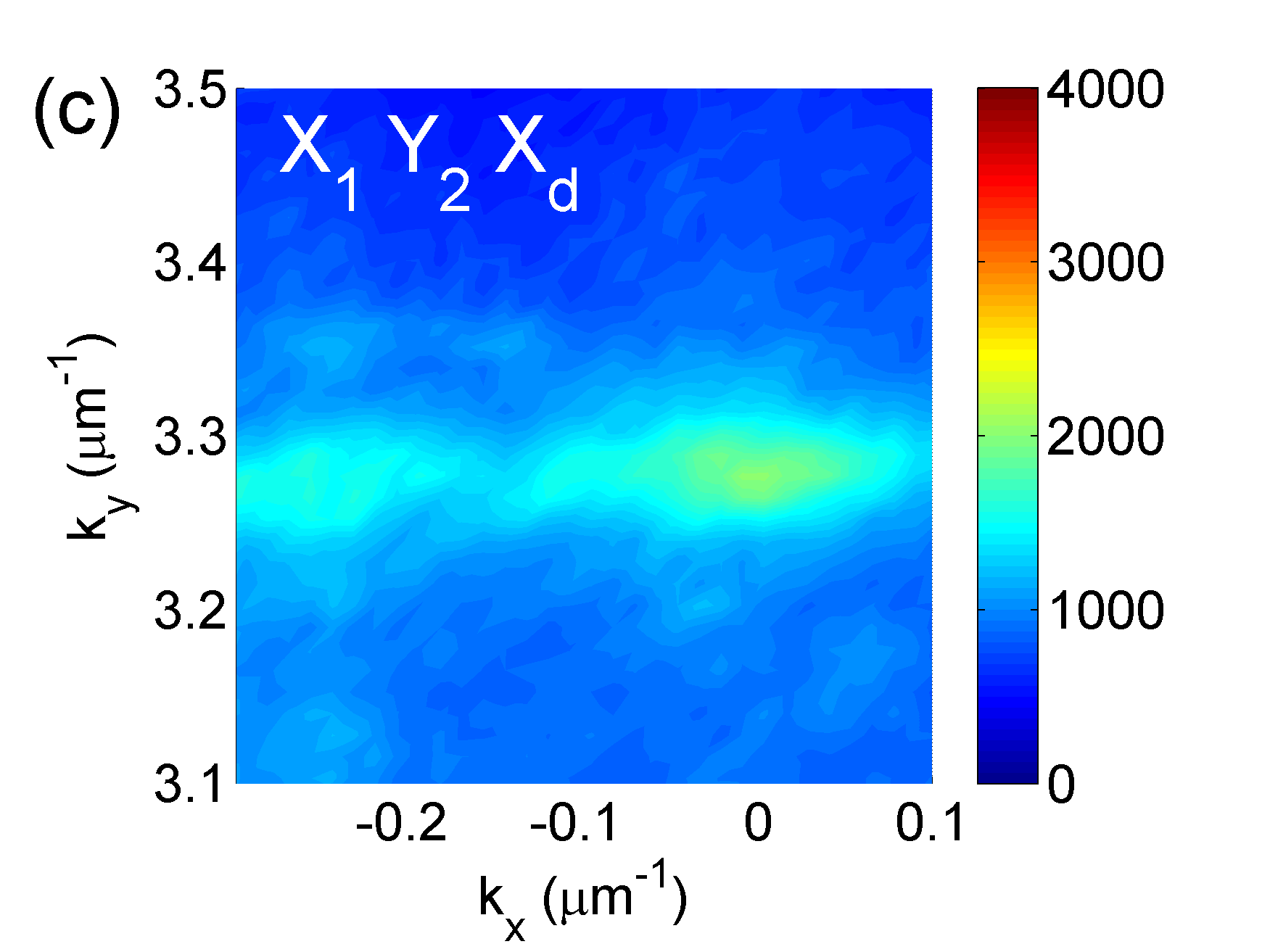} \includegraphics[height=3cm]{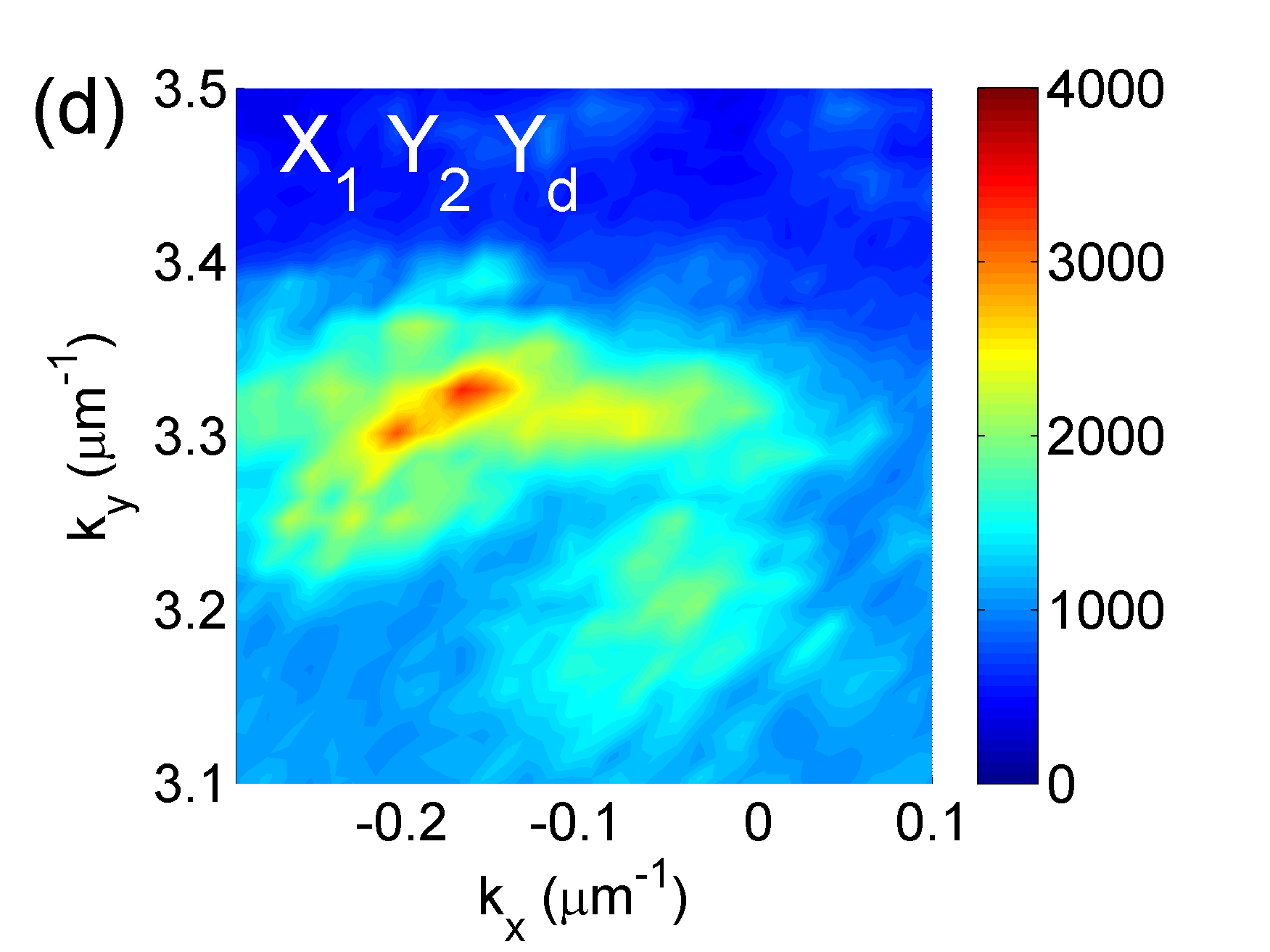}
\includegraphics[height=3cm]{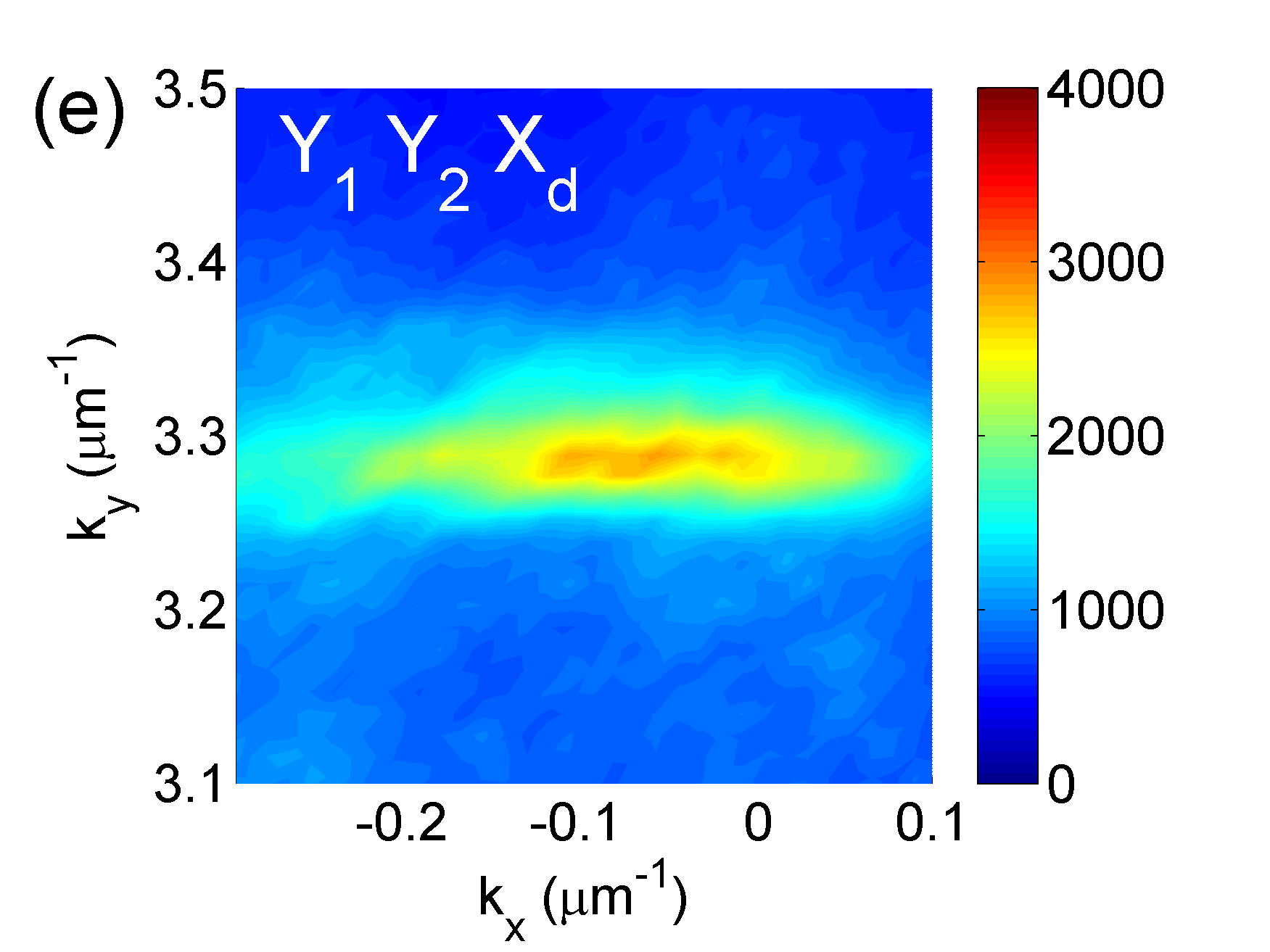} \includegraphics[height=3cm]{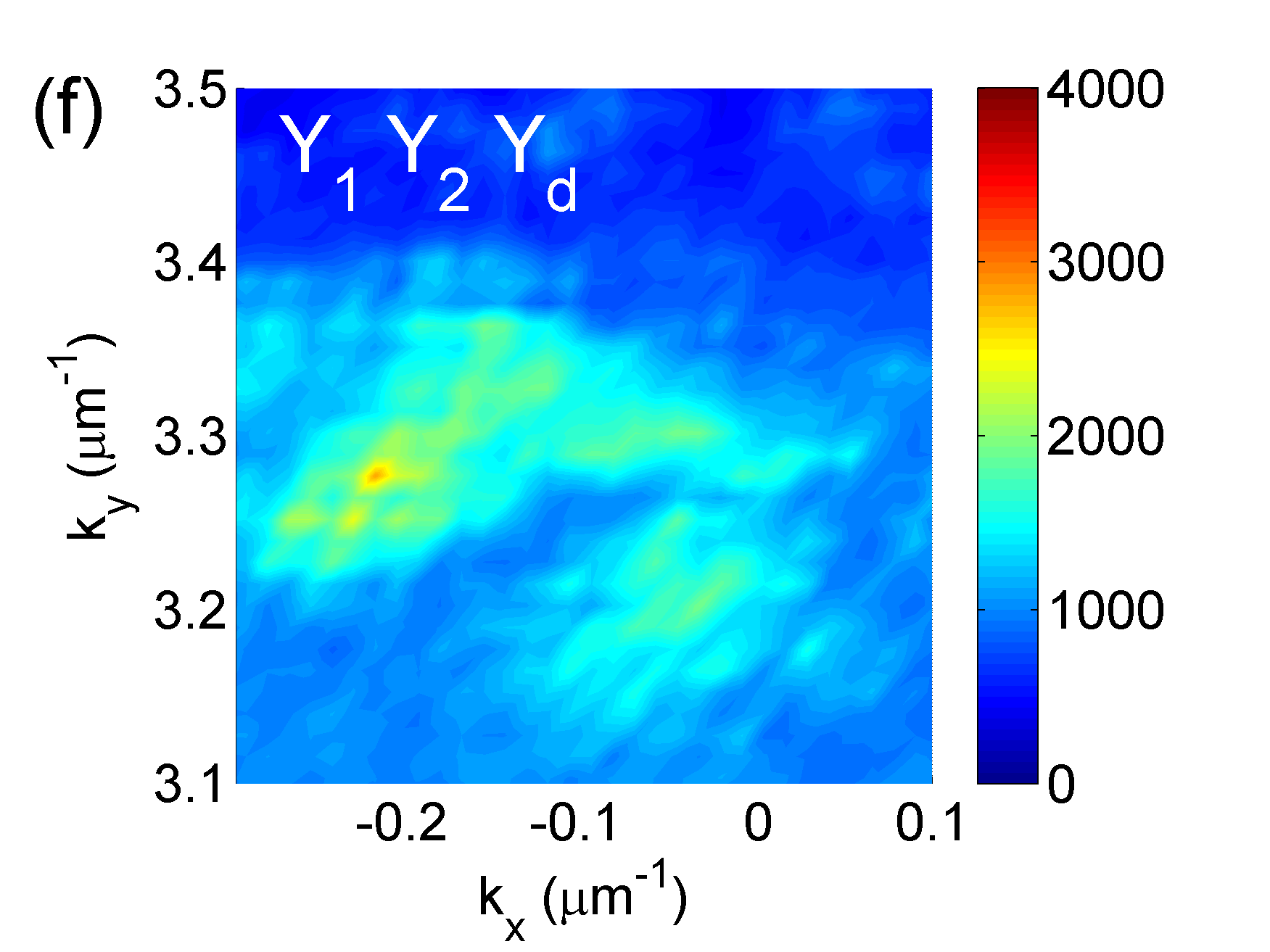} 
\caption{
\textcolor{black}{Zoom into the k-space region of interest (enclosed by the red rectangle in Fig.~ \ref{fig:double_probeexp_scheme})   the SOFWM. Results are shown for the following excitation/detection configurations. The left column (a,c,e) shows the detection in copolarized ($\mathrm{X_d}$), the right  column (b,d,f) in cross-polarized ($\mathrm{Y_d}$) detection. Here, either both probes are copolarized (a, b), one probe is co- and the second is cross-polarized (c, d), or both probes are cross-polarized (e, f).  
}}
\label{fig:double_probeexp}
\end{figure}

The experimental results measured for different polarization configurations of two linearly polarized probes are shown in Figs.~\ref{fig:double_probeexp_scheme} and \ref{fig:double_probeexp}: In Fig.~\ref{fig:double_probeexp_scheme} the measurement configuration in the full momentum space plane is shown, the panels in Fig.~\ref{fig:double_probeexp} only show a zoom into the k-space range of relevance for the SOFWM.

To analyze these results theoretically, it is useful to transform Eq.~(\ref{eq:GPE_p}) into the linear polarization basis for the excitonic component:
\begin{widetext}
\begin{align}
\mathrm{i} \hbar \dot{p}_i^{X \atop Y} & = \left( \mathbb{H}^{x} - \mathrm{i} \gamma_x \right) p_i^{X \atop Y} - \Omega_x E_i^{X \atop Y} + {1 \over 2 } \alpha_\mathrm{PSF} \Omega_x \left( \vert p_i^{{X \atop Y}} \vert^2 + \vert p_i^{{Y \atop X} } \vert^2  \right) E_i^{X \atop Y}
+ {1 \over 2 } \alpha_\mathrm{PSF} \Omega_x \left( p_i^{{Y \atop X}*} p_i^{X \atop Y}  -  p_i^{{X \atop Y}*} p_i^{Y \atop X}   \right) E_i^{Y \atop X} \nonumber \\
& + {1 \over 2} \left( T^{++} + T^{+-} \right) \vert p_i^{{X \atop Y} } \vert^2  p_i^{X \atop Y}  - {1 \over 2}  \left( T^{++} - T^{+-} \right) p_i^{{X \atop Y} *}   \left( p_i^{Y \atop X} \right)^2  + T^{++} \vert p_i^{{Y \atop X}} \vert^2
 p_i^{X \atop Y}.
\label{eq:double_probe}
\end{align}
\end{widetext}

%
%

Based on this expression, the SOFWM can be analyzed analytically. In our setup, the pump-induced polaritons are vertically polarized. Then, considering the nonlinear terms for the probe polarization configurations used in the experiment in Fig.~\ref{fig:double_probeexp}, we find the following selection rules: (i) If both off-axis (probe) fields are co-polarized to the pump, then the SOFWM is co-polarized as well (X$_1$X$_2$X$_\mathrm{d}$). This explains the stronger signal observed in Fig.~\ref{fig:double_probeexp}a where the detection is in the X channel. Only a relatively weak signal is detected in the Y channel  (Fig.~\ref{fig:double_probeexp}b). (ii) One probe field is co-, the other cross-polarized to the pump. Then, the SOFWM is cross-polarized. This means the main signal is in the X$_1$Y$_2$Y$_\mathrm{d}$ configuration shown in Fig.~\ref{fig:double_probeexp} d. Only weak SOFWM is measured in X$_1$Y$_2$X$_\mathrm{d}$ configuration  (Fig.~\ref{fig:double_probeexp}c). (iii) Both off-axis fields are cross-polarized to the pump. Then, the main SOFWM signal is measured in the Y$_1$Y$_2$X$_\mathrm{d}$ configuration in Fig.~\ref{fig:double_probeexp}e compared to Y$_1$Y$_2$Y$_\mathrm{d}$ configuration in Fig.~\ref{fig:double_probeexp}f.

\begin{figure}
\includegraphics[height=3cm]{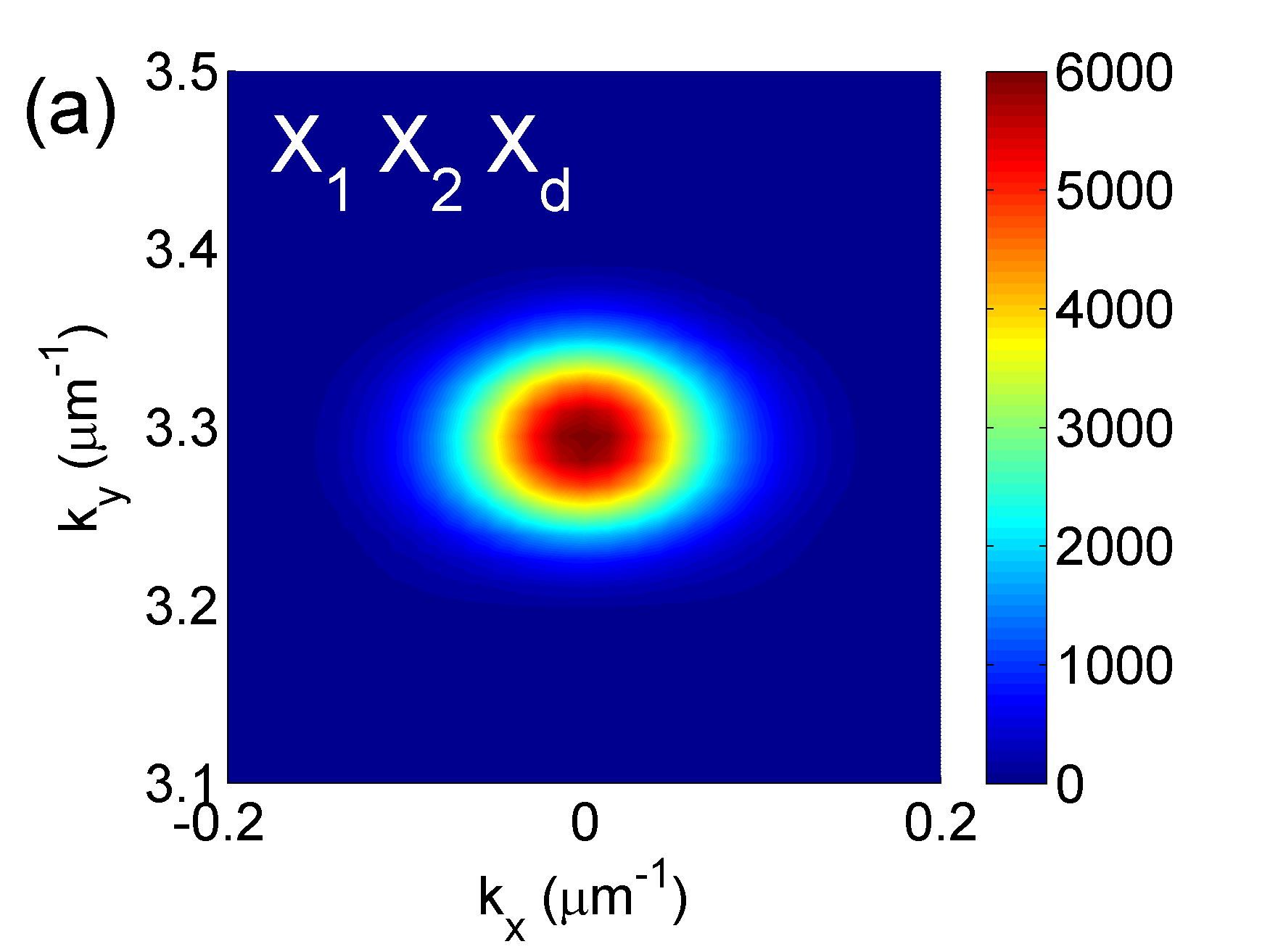} \includegraphics[height=3cm]{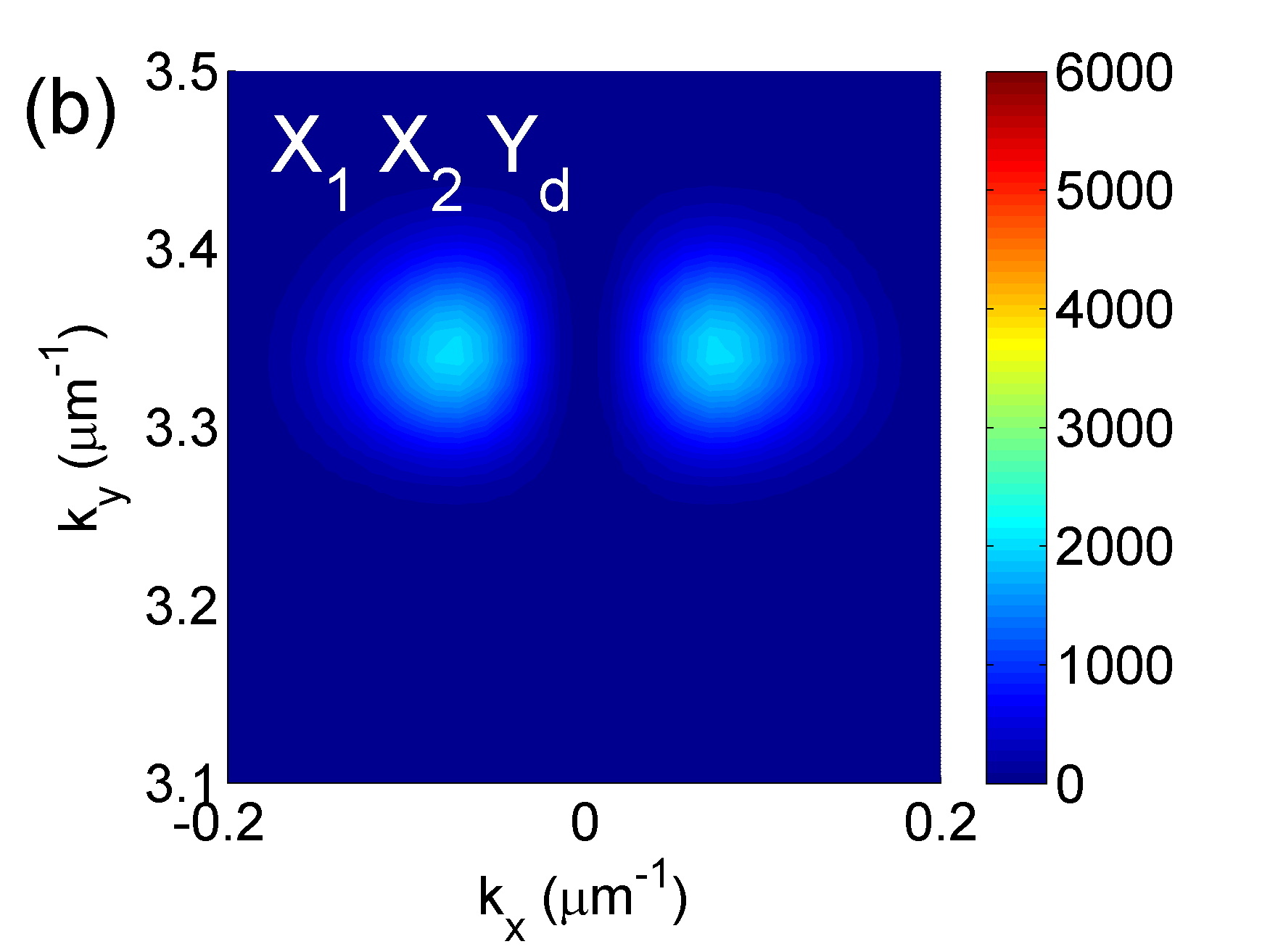}
\includegraphics[height=3cm]{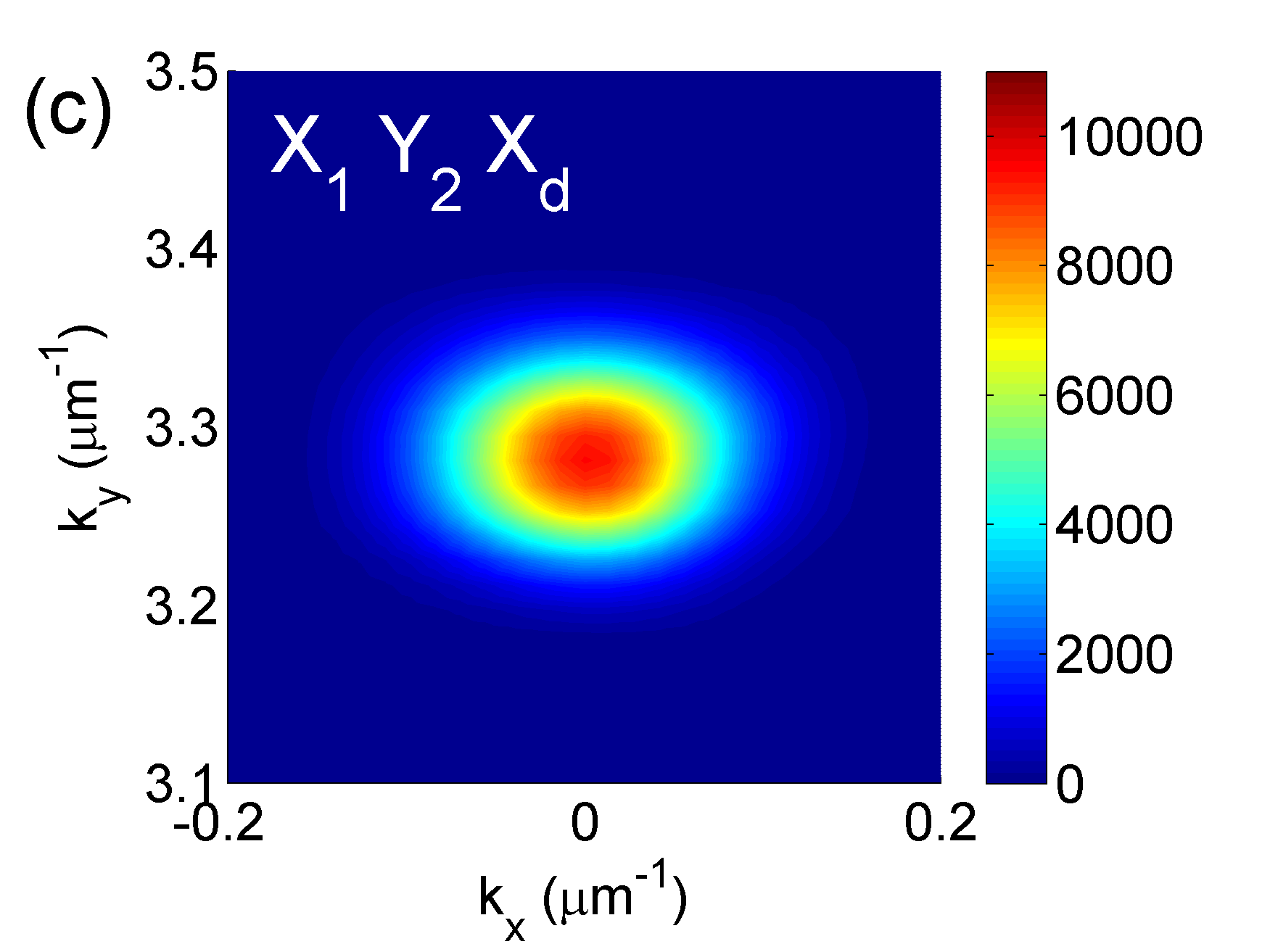} \includegraphics[height=3cm]{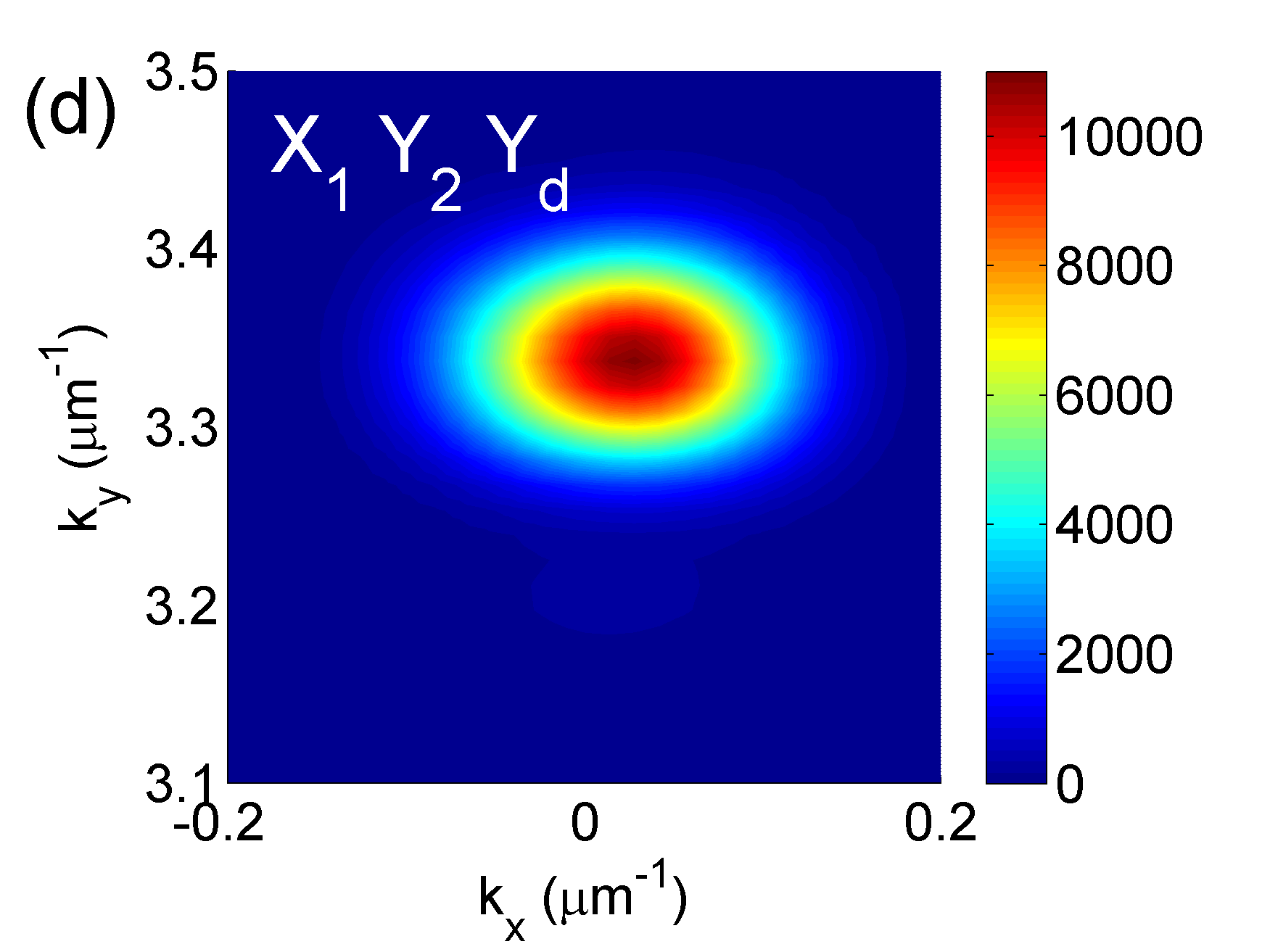}
\includegraphics[height=3cm]{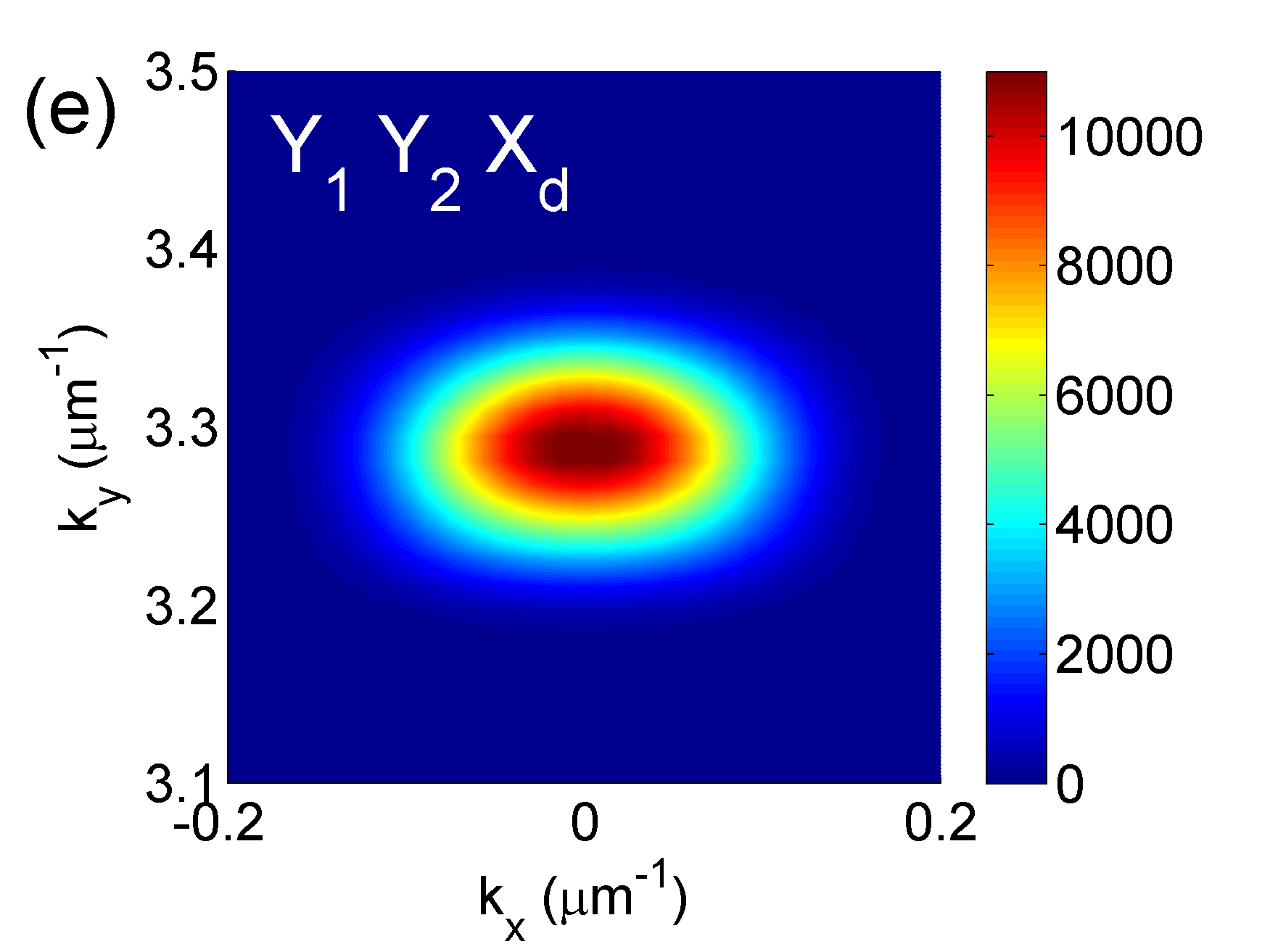} \includegraphics[height=3cm]{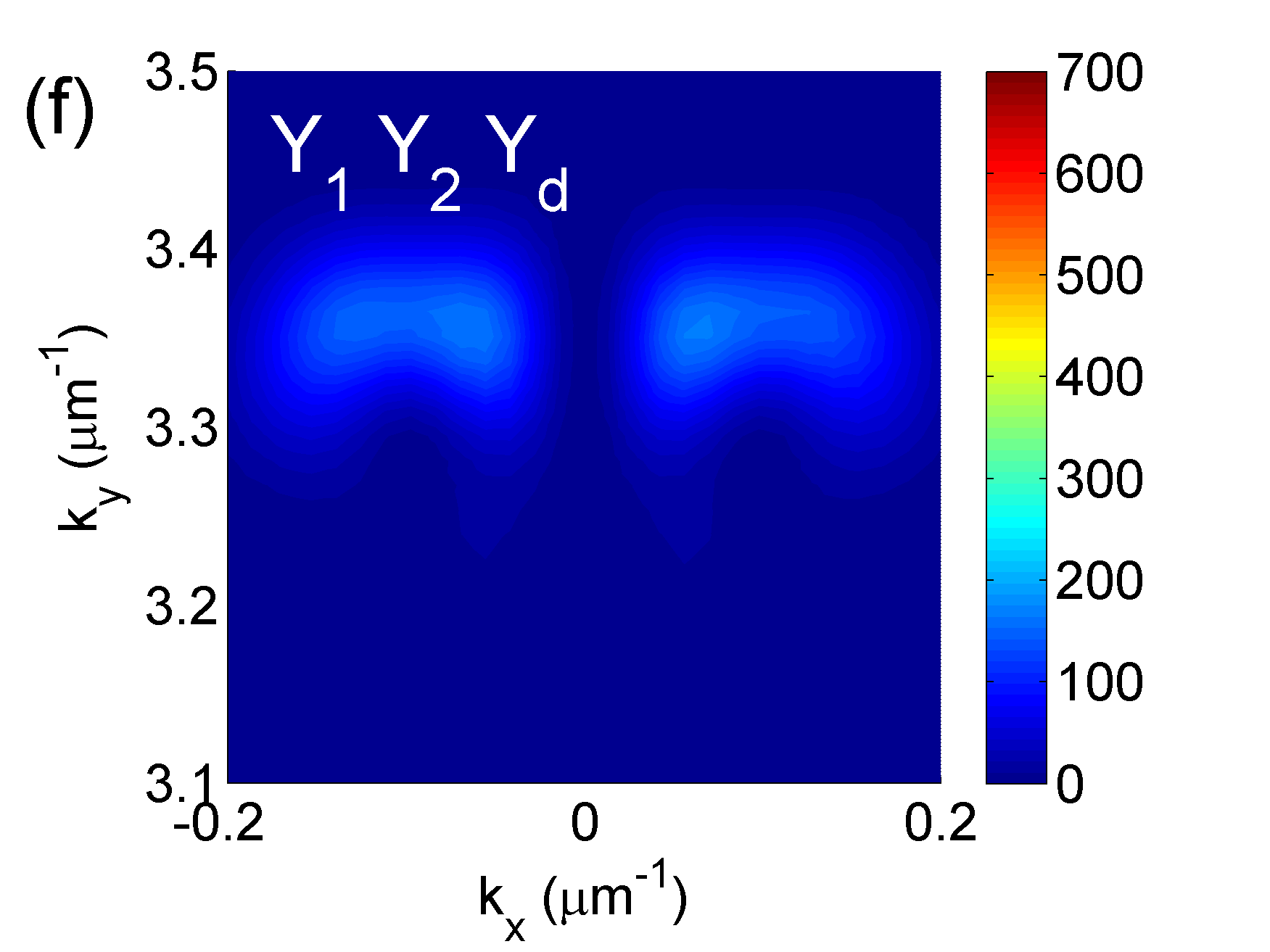}
\caption{
\textcolor{black}{Simulation results with a pump and two probe beams to analyze second-order four-wave mixing processes for the same excitation/detection configurations as in Fig.~\ref{fig:double_probeexp}. A zoom into the k-space region of interest for SOFWM is shown.}}
\label{fig:double_probesim}
\end{figure}

The selection rules discussed above have only been derived from the excitonic part (giving rise to the nonlinearities) of the polariton field. However, due to the presence of TE-TM splitting these selection rules do not strictly hold but are alleviated such that weaker SOFWM signals are also found in the ``forbidden" polarization channels. In Fig.~\ref{fig:double_probesim} we show simulations based on Eqs.~(\ref{eq:GPE_e}) and (\ref{eq:GPE_p}) corresponding to the polarization configurations studied in Fig.~\ref{fig:double_probesim}.

As in the experiments, we find the SOFWM signal predominantly in the expected polarization channels, but TE-TM splitting yields smaller SOFWM signatures also in the ``forbidden" channels. We note that in the calculations additional details are resolved:
Since the probe light is spread out in k-space, it does not map sharply the TE eigenmode only. As a consequence, a signal occurs in X$_1$X$_2$Y$_\mathrm{d}$ (Fig.~\ref{fig:double_probesim} b) and Y$_1$Y$_2$Y$_\mathrm{d}$-configuration (Fig.~\ref{fig:double_probesim} f) also in the TM eigenmode at $k = 3.3 \, \mu\mathrm{m}^{-1}$ which is then further split into two peaks due to TE-TM splitting.

In summary, also the selection rules observed for the second-order FWM can be satisfactorily explained based on the $\chi^{(3)}$ nonlinearity included in our theoretical approach evaluated in higher order. While this may not exclude possible contributions from many-particle correlations beyond two-exciton interactions (which are neglected in the theoretical analysis used here), for the excitation regime studied it is indicative of the dominant role of processes included in our approach.
%
%
%
\section{Conclusions}
\label{sec:conclusion} 
We have presented a detailed analysis of the polarization dependence of four-wave mixing processes in a spinor fluid of microcavity polaritons. We report results on first and second order processes and find good agreement between experimental data, numerical simulations, and analytical calculations. 

In the first order process, we find that the FOFWM signal is largest when probe and detection are cross-polarized to the pump, which is consistent with the fact that exciton-exciton interaction favors scattering of pump polaritons into cross-polarized states. We also see clear signatures of the TE-TM splitting, which yields different elastic circles and hence different k-magnitudes for the scattered FOFWM signals. Taken together, these two effects explain the four-fold symmetry seen in the data shown in Fig.~\ref{fig:single_probe_results}. Another clear signature showing the influence of TE-TM splitting is the presence of a FWM signal in the detection channel that is cross-polarized to the probe. 

In the second order process, where we use two probes separated by 120$^\circ$ on the elastic circle, we find that there is indeed a SOFWM signal at the expected location on the elastic circle, see Fig.~\ref{fig:pump_probe_scheme}b. This shows clearly that wave mixing processes favoring 60$^\circ$ scattering on the elastic circle are present in our system. A detailed analysis of the polarization channels yields good agreement between experiment and theory. Since our theory includes interactions only up to two-exciton correlations (up to four-particle interactions in the electron-hole picture), we conclude that nonlinearities beyond this level are not necessary to understand the experimentally observed data, even though two-probe experiments are in principle sensitive to higher-order correlations.

The present study has been done on a double-cavity system which brings the possibility to observe the formation of polariton patterns. The pattern formation is crucially determined by the four-wave mixing processes studied here. The present study reveals the central role played by TE-TM splitting and spin-dependent parametric scattering in pattern formation and orientation. This work, by extending our understanding and mastering of those mechanisms, pave the way to the development of improved application oriented concepts such as all-optical switches based on semiconductor microcavities. 

\section{Acknowledgements}
The Paderborn group acknowledges financial support from the DFG (SCHU 1980/5-1, GRK 1464, and TRR 142) and a grant for computing time at $\mathrm{PC^2}$ Paderborn Center for Parallel Computing. Stefan Schumacher further acknowledges support through the Heisenberg program of the DFG. The Arizona group was supported by the National Science Foundation (NSF) under grant ECCS-1406673, and by TRIF Photonics.
\section{Appendix A - Details of the theoretical analysis}
\label{sec:AppendixA}
In section \ref{sec:sec:LSA}, the dynamic of probe and FWM photons and excitons is coupled by a 16x16 matrix $\mathbb{M}$ in Eq.~(\ref{eq:LSA}). The following 4x4 block matrices were introduced to express $\mathbb{M}$:
\begin{equation}
\mathbb{M}^i_{EE} = \left(\begin{array}{cccc} \Delta \hbar \omega^c_k & \epsilon_k^{+} & 0 & 0 \\
\epsilon_k^{-} & \Delta \hbar \omega^c_k & 0 & 0 \\
0 & 0 & - \Delta \hbar \omega_k^{c*} & - \epsilon_k^{+*}\\
0 & 0 & - \epsilon_k^{-*} & - \Delta \hbar \omega_k^{c*} \end{array}\right),
\end{equation}
couples the photonic polarization components in each cavity. $\Delta \hbar \omega_k^c = \hbar \omega^c_k + {{\hbar^2} \over {4}} \left( {{1}\over{m_\mathrm{TM}}} + {{1} \over {m_\mathrm{TE}}} \right) k^2 - \hbar \omega_\mathrm{pump} - \mathrm{i} \gamma_c$ denotes the frequency detuning in respect to the photonic energy and the off-diagonal coupling $\epsilon_k^{\pm} = {{\hbar^2} \over {4}} \left( {{1}\over{m_\mathrm{TM}}} - {{1} \over {m_\mathrm{TE}}} \right) \left(  k_x  \mp \mathrm{i} k_y \right)^2$ due to TE-TM splitting. The matrix
\begin{equation}
\mathbb{M}^i_{pp} = \left(\begin{array}{cccc} \Delta \epsilon^{x +}_{i} & U_i^{+} & V_i^{+} & W_i^{+} \\
U_i^{-} & \Delta \epsilon^{x -}_i & W_i^{-} & V_i^{-} \\
-V_i^{+*} & -W_i^{+*} & -\Delta \epsilon^{x+*}_i & -U_i^{+*}\\
-W_i^{-*} & -V_i^{-*} & -U_i^{-*} & -\Delta \epsilon^{x-*}_i \end{array}\right)
\end{equation}
couples the excitonic components in each cavity $i$. The diagonal terms $\Delta \epsilon^{x\pm}_i = {\epsilon^x_0} + \alpha_\mathrm{PSF} \Omega_x p^{\pm*}_{0,i} E^{\pm}_{0,i} + 2 T^{++}\vert p^{\pm}_{0,i} \vert^2 +  T^{+-}\vert p^{\mp}_{0,i} \vert^2 - \hbar \omega_\mathrm{pump} - \mathrm{i} \gamma_x$ are blue-shifted frequency-detunings in respect to the exciton energy
and $U_i^{\pm} = T^{+-} p^{\mp *}_{0,i}  p^{\pm}_{0,i}$, $V_i^{\pm} = \alpha_\mathrm{PSF} \Omega_x p^{\pm}_{0,i} E^{\pm}_{0,i} + T^{++} \left( p^{\pm}_{0,i} \right)^2$ and
$W_i^{\pm} = T^{+-} p^{\mp}_{0,i}  p^{\pm}_{0,i}$ are nonlinear constants including the exciton density at $k=0$.
{\color{black} We note that also the pump-induced exciton fields are not the same in the two cavities and reflect the partial asymmetry of the optically pumped mode with $E_1 = -1.53 E_2$ and $p^{\pm}_{0,1} = -1.5 p^{\pm}_{0,2}$.}
The photon-exciton coupling in each cavity is given by
\begin{equation} \mathbb{M}^i_{Ep} = \left(\begin{array}{cccc} -\Omega_X & 0 & 0 & 0 \\
 0 & -\Omega_X & 0 & 0 \\
 0 & 0 & \Omega_X & 0 \\
 0 & 0 & 0 & \Omega_X  \end{array}\right)\end{equation}
and
\begin{equation} \mathbb{M}^i_{pE} = \left(\begin{array}{cccc} -\tilde{\Omega}_{X,i}^{+} & 0 & 0 & 0 \\
 0 & -\tilde{\Omega}_{X,i}^{-} & 0 & 0 \\
 0 & 0 & \tilde{\Omega}_{X,i}^{+} & 0 \\ 
 0 & 0 & 0 & \tilde{\Omega}_{X,i}^{-}  \end{array}\right)\,,\end{equation}
where the phase-space filling $\tilde{\Omega}_{X,i}^{\pm} = \Omega_X \left( 1 - \alpha_\mathrm{PSF} \vert p^{\pm}_{0,i} \vert^2  \right)$ is included in the coupling constant. Both cavities are coupled by
\begin{equation} \mathbb{M}_{12} = \left(\begin{array}{cccc} -\Omega_C & 0 & 0 & 0 \\
 0 & -\Omega_C & 0 & 0 \\
 0 & 0 & \Omega_C & 0 \\
 0 & 0 & 0 & \Omega_C  \end{array}\right)\end{equation}
through the photonic component with $\mathbb{M}_{12} = \mathbb{M}_{21}$.
%
%
%
\section{Appendix B - Simplified two-component equation}
\label{sec:AppendixB}
To give a more intuitive picture of the polarization-dependent processes in the polariton fluid created by the pump beam, it is instructive to only discuss a simplified equation. Phenomenologically, the dynamics of the polariton field in the two polarization channels of the lowest polariton branch can be governed by:
\begin{eqnarray}
\mathrm{i} \hbar\frac{\partial}{\partial t}\Psi^{\pm} & = & \left(\mathbb{H}_{p}-\mathrm{i}\gamma_p \right)\Psi^{\pm}+\mathbb{H}_p^{\pm}\Psi^{\mp} +\alpha^{++}\vert\Psi^{\pm}\vert^{2}\Psi^{\pm} \nonumber \\ & & +\alpha^{+-}\vert\Psi^{\mp}\vert^{2}\Psi^{\pm}+\Psi_\mathrm{pump}^{\pm}\,.
\label{eq:simple_GPE}
\end{eqnarray}
Here, $\psi^\pm$ denotes the polariton field amplitude on a parabolic polariton dispersion. In analogy to the scenario studied above, the pump source is at $k=0$ and tuned above the bottom of the polariton dispersion such that stimulated and resonant scattering onto the polariton dispersion can occur. This simplified equation contains all the details needed to describe the polarization-dependent effects observed above qualitatively. The general conclusions drawn qualitatively agree with the LSA for the full model. In analogy to the full model discussed above, the diagonal and off-diagonal dispersion (TE-TM splitting) is included by $\mathbb{H}_{p}$ and $\mathbb{H}_p^{\pm}$, respectively. The polariton decay is given by $\gamma_p$. The interactions between polaritons \cite{lecomte.14} is given by $\alpha^{++}$ for cocirularly and $\alpha^{+-}$ for countercircularly polarized polaritons, corresponding to  $T^{++}$ and $T^{+-}$ in Eq.~\ref{eq:GPE_p}. $\Psi_\mathrm{pump}^{\pm}$ is the pump source. The parameters can be extracted from the full equations via the Hopfield-coefficients \cite{Saito2013}.

In this simplified model, we can apply the LSA like in sec. \ref{sec:sec:LSA} {\color{black} based on Eq. (\ref{eq:simple_GPE})}. Instead by a 16x16 matrix, the initial dynamics of off-axis polaritons can be described by the following equation including only a 4x4 matrix:
\begin{widetext}
\begin{align}
\frac{\partial}{\partial t} & \left(\begin{array}{l}
\Psi_{p}^{X}\\
\Psi_{f}^{X*}\\
\Psi_{p}^{Y}\\
\Psi_{f}^{Y*}\end{array}\right) \nonumber \\
& =\frac{-\mathrm{i}}{\hbar}\left(\begin{array}{cccc}
\Delta\epsilon^X_{k}+\Gamma_{k}\cos(2\phi) & \frac{1}{2}\left(\alpha^{++}+\alpha^{+-}\right)\Psi_{0}^{X 2} & \Gamma_{k}\sin(2\phi) & 0\\
-\frac{1}{2}\left(\alpha^{++}+\alpha^{+-}\right)\Psi_{0}^{X*2} & -\Delta\epsilon^{X*}_{k}-\Gamma_{k}\cos(2\phi) & 0 & -\Gamma_{k}\sin(2\phi)\\
\Gamma_{k}\sin(2\phi) & 0 & \Delta\epsilon^Y_{k}-\Gamma_{k}\cos(2\phi) & -\frac{1}{2}\left(\alpha^{++}-\alpha^{+-}\right)\Psi_{0}^{X 2}\\
0 & -\Gamma_{k}\sin(2\phi) & \frac{1}{2}\left(\alpha^{++}-\alpha^{+-}\right)\Psi_{0}^{X * 2} & -\Delta\epsilon_{k}^{Y*}+\Gamma_{k}\cos(2\phi)\end{array}\right)\left(\begin{array}{l}
\Psi_{p}^{X}\\
\Psi_{f}^{X*}\\
\Psi_{p}^{Y}\\
\Psi_{f}^{Y*}\end{array}\right)\,.
\label{eq:simple_LSA}
\end{align}
\end{widetext}

To derive Eq.~\ref{eq:simple_LSA} also a transformation into a linearly polarized basis was used. In the basis chosen, the polarization components are either parallel or perpendicular to the pump polarization plane (X or Y, respectively). The vector $\left(\Psi_{p}^{X},\Psi_{f}^{X*},\Psi_{p}^{Y},\Psi_{f}^{Y*}\right)^{\mathrm{T}}$ contains probe and FWM for each polarization state, as discussed in sec. \ref{sec:sec:LSA}. In the matrix,
$\Psi_{0}^{X}$ denotes the amplitude of the homogenous polariton field solution, $\Delta \epsilon^X_{k}=\epsilon^p_k +\left(\alpha^{++}+\alpha^{+-}\right) \vert\Psi_{0}^{X}\vert^{2}-\hbar\omega_\mathrm{pump}-\mathrm{i}\gamma_p$ and $\Delta\epsilon^Y_{k}=\epsilon^p_k +\alpha^{++}\vert\Psi_{0}^{X}\vert^{2} -\hbar\omega_\mathrm{pump}-\mathrm{i}\gamma_p$  are energy detuning ($\epsilon^p_k$ denotes the polariton dispersion) and $\Gamma_{k}=\frac{\hbar^{2}k^{2}}{4}\left(\frac{1}{m_{\mathrm{TM},p}}-\frac{1}{m_{\mathrm{TE},p}}\right)$ is the strength of the TE-TM splitting ($m_{\mathrm{TM},p}$ and $m_{\mathrm{TE},p}$ denote the effective masses for TM and TE mode, respectively).

For vanishing TE-TM splitting ($\Gamma_k = 0$) and $\alpha^{+-} = 0$ as presented in Fig.~\ref{fig:LSA}a and d, the matrix $\mathbb{M}$, in Eq.~\ref{eq:simple_LSA} becomes block-diagonal and independent of $\phi$. Solving the eigenvalues exhibits the amplification of off-axis polaritons by scattered pump polaritons is azimuthally symmetric and marks a circle in the momentum space with a well-defined radius $k_{res}$.
Including TE-TM splitting, $\mathbb{M}$ is $\phi$-dependent and only block-diagonal in the cases $\phi=0,\frac{n\pi}{2}$ (in these cases $\sin (2\phi)$ vanishes).
Instead of $\Delta \epsilon^{X}_k = \Delta \epsilon^{Y}_k$ in both blocks, now $\Delta \epsilon^X_k + \Gamma_k$ and $\Delta \epsilon^Y_k - \Gamma_k$ enters in the resonance condition.
Therefore, there is a splitting of the circle with different $k_{res}$ in the vicinity of $\phi=\frac{n\pi}{2}$, forming two crescents (Fig.~\ref{fig:LSA}b and e). Depending on the sign of $\cos (2\phi)$, the corresponding eigenmodes of the upper crescent are X (Y)-polarized for $\phi = \frac{\pi}{2}, \frac{3\pi}{2}$ ($\phi = 0, \pi$), the eigenmodes for the lower crescent Y (X)-polarized for $\phi = \frac{\pi}{2}, \frac{3\pi}{2}$ ($\phi = 0, \pi$).
For a vanishing $\alpha^{+-}$, the amplification by pump polaritons is balanced for the X- and Y-polarization channel (the effective decay has the same magnitude on both crescents). However, for an $\alpha^{+-} \neq 0$ the eigenvalues of each block matrix of $\mathbb{M}$ are not identical anymore. For an $\alpha^{+-} < 0$, the ``effective" nonlinearity for the Y-channel ($\alpha^{++}-\alpha^{+-}$) is greater than for the X-channel ($\alpha^{++}+\alpha^{+-}$): The consequence is a preffered scattering of pump polaritons into the cross-polarized channel (Fig. \ref{fig:LSA}c and f). The fourfold symmetry of Fig. \ref{fig:LSA}b is broken.
%
%
%
%
\bibliography{./allref,./stefan-literature,./pattern}

\bibliographystyle{spiebib}

\end{document}